\pdfoutput=1

\documentclass[preprint, 3p, twocolumn, times,sort]{elsarticle}

\usepackage{pgfplots}
\pgfplotsset{width=7cm,compat=1.3}

\usepackage{balance}
\usepackage{booktabs}
 \usepackage{siunitx}
 \usepackage{flushend}

\usepackage{url}

\usepackage{booktabs}

\usepackage{tikz}
\usepackage{color}
\usetikzlibrary{chains,positioning,arrows}

\usepackage{subfloat,subfig,float,graphicx,grffile,verbatim,url}
\usepackage{multicol,multirow}
\usepackage{algorithm}
\usepackage[noend]{algorithmic}
\usepackage{flushend}
\usepackage[colorlinks,bookmarksopen,bookmarksnumbered,citecolor=red,urlcolor=red,draft]{hyperref}
 \usepackage{siunitx}
\DeclareSIUnit \GHz {\giga{}\hertz{}}
\DeclareSIUnit \byte {B}
\DeclareSIUnit \MB {MB}
\DeclareSIUnit \kB {kB}
\DeclareSIUnit \GB {GB}

\usepackage{listings}
\graphicspath{{./gnuplot/}}

\usepackage{todonotes}

\journal{Information Systems}

\begin{document}
\begin{frontmatter}

\title{Upscaledb: Efficient  Integer-Key Compression in a Key-Value Store  using SIMD~Instructions}

\author[daniel]{Daniel Lemire\corref{cor1}}
\ead{lemire@gmail.com}
\address[daniel]{LICEF Research Center, TELUQ University of Quebec, Canada}

\author[christoph]{Christoph Rupp}
\ead{chris@crupp.de}
\address[christoph]{Upscaledb, Munich, Germany}

 \cortext[cor1]{Corresponding author. Tel.: 00+1+514 843-2015 \#2835; fax: 00+1+800 665-4333.}

\begin{abstract}
Compression can sometimes improve performance by making more of the data available
to the processors faster.
We consider the compression of integer keys in a B+-tree index. 
For this purpose,
systems such as IBM DB2 use variable-byte compression over differentially coded keys. 
We revisit this problem with various  compression alternatives such as Google's VarIntGB, Binary Packing and Frame-of-Reference. In all cases, we describe algorithms that can operate directly on compressed data. 
Many of our alternatives exploit the single-instruction-multiple-data (SIMD) instructions  supported by modern CPUs. 
We evaluate our techniques in a database
environment provided by Upscaledb, a production-quality key-value database.
Our  best techniques are SIMD accelerated:
they  simultaneously reduce memory usage
while improving single-threaded speeds. 
In particular, a differentially coded SIMD binary-packing techniques (BP128) can offer a superior
 query speed (e.g., \SI{40}{\percent} better than
an uncompressed database) while providing the best compression (e.g., by a factor
of ten).
For analytic workloads, our fast compression techniques
offer compelling benefits. Our software is available as open source.
\end{abstract}


\end{frontmatter}

\lstdefinestyle{customc}{%
  belowcaptionskip=1\baselineskip,
  breaklines=true,
  xleftmargin=\parindent,
  language=C,
  showstringspaces=false,
  basicstyle=\small\ttfamily,
  keywordstyle=\bfseries\color{green!40!black},
  commentstyle=\itshape\color{purple!40!black},
  identifierstyle=\bfseries\color{black},
  stringstyle=\color{orange},
   morekeywords={uint64_t,uint32_t,__m256i,__m128i},
}

\lstset{escapechar=@,style=customc}

\section{Introduction}

The B-tree and its variations (such as the B+-tree) have been ubiquitous in computing since the 1970s~\cite{Bayer:1972:OML:2697402.2697450}. Almost all relational
database engines use  data structures resembling the B-tree. Moreover, 
many applications rely on an embedded database to store key-value pairs in B-trees. There are many popular choices today including Berkeley~DB~\cite{Olson:1999:BD:1268708.1268751}, Kyoto Cabinet and Upscaledb.

Compressing B-trees can help reduce storage and it may even accelerate some queries by easing the data transfer bottleneck. Indeed, many operations in  modern databases leave the CPU idle, waiting for data to arrive either from RAM or from the disk. 
Applications where updates are infrequent and queries have low selectivity (i.e., analytic workloads)  are especially likely to benefit from compression. 

Short fixed-length keys are well suited for advanced optimization.
Of particular interest is the case where keys are integer values (e.g., 32-bit integers). They are relevant when the key is an identifier (e.g., IP address, user ID, row ID and so forth).

To get substantial performance benefits out of compression, we must ensure that decompression is fast enough.
For this purpose,  IBM DB2 uses variable-byte compression and differential coding~\cite{db2luw2009}. In a differential-coding model, instead of storing the integer keys themselves ($x_1, x_2, \ldots$), we store the successive differences ($x_1-0, x_2-x_1, x_3-x_2, \ldots$). These differences are typically small when the values are maintained in  sorted order. Small integers can be compressed quickly (see \S~\ref{sec:codecs}). The compression technique used by IBM~DB2 is well established, fast and it provides reasonable compression ratios.  

However, there is a wide range of integer compression techniques, and many of them have been modified to fully benefit from modern processors.  For example, commodity processors (e.g., ARM, POWER, Intel, AMD) provide instructions operating on wide 128-bit registers (called \emph{XMM registers} on Intel platforms). 
These wide registers can be used
to operate on many values at once (e.g., four 32-bit integers). So we can add or subtract four pairs of 32-bit integers using a single instruction.  We qualify these instructions as being  single-instruction-multiple-data (SIMD). They are  helpful when compressing arrays of integer values~\cite{Zhao:2015:GSA:2737814.2735629,LemireBoytsov2013decoding}.

Can optimized compression techniques accelerate a key-value database
system? To find out, we implemented a set of state-of-the-art compression techniques in an established open-source embedded database engine (Upscaledb) based on a B+-tree. We also implemented accompanying algorithms over compressed data to support 
common
operations.
 We provide an experimental comparison.  To our knowledge, such a comparison has never been attempted, especially not one that includes SIMD-accelerated schemes. 

Unsurprisingly, we show that we can compress the keys of the B+-tree
to a tenth of the original size. This ensures that more data can be
stored in RAM\@.
Moreover, our results indicate that compression improves real-world performance and that the gains can be substantial (e.g., \SI{40}{\percent}). However, we find that only our fastest schemes, those based on SIMD instructions, accelerate B+-tree performance reliably. 
Our work confirms earlier findings stressing the benefits of designing compression algorithms to leverage SIMD instructions~\cite{SPE:SPE2326,Zhao:2015:GSA:2737814.2735629}.

\section{Codecs}
\label{sec:codecs}

We are interested in compressing 32-bit unsigned integers.
For most of our compression algorithms, integers are differentially coded prior to compression so
that most of them are small. That is, starting from an array of
integers $x_1, x_2, \ldots$, we compress the integers
$x_1, x_2-x_1, x_3-x_2, \ldots$
There are many suitable integer-compression schemes: we selected some likely
to offer good performance.

For recent reviews on the topic, see Lemire and Boytsov~\cite{LemireBoytsov2013decoding} or
Zhao et al.~\cite{Zhao:2015:GSA:2737814.2735629}.

\paragraph{Differential coding}
During decoding, given the differences 
$\delta_1=x_1, \delta_2=x_2-x_1, \delta_3=x_3-x_2, \ldots$,
we need to reconstruct $x_1, x_2, x_3, \ldots$ This operation requires the 
computation of a prefix sum 
($\delta_1, \delta_1 + \delta_2, \delta_1+\delta_2+\delta_3, \ldots$).
To avoid unnecessary data operations, we integrate the decompression
and the prefix sum computation. That is, we output the decoded values
all at once: we do not first output the difference
$\delta_1, \delta_2, \delta_3, \ldots$ and then the reconstructed
decoded values.
For SIMD-based schemes, the computation of the prefix sum can be vectorized
as follows. We process the data in registers of four~32-bit integers (which call vectors).
\begin{enumerate}
\item Shift the vector by two integers
$(\delta_i,\delta_{i+1},\delta_{i+2},\delta_{i+3}) \to  (0,0,\delta_{i},\delta_{i+1})$.
\item Add the original delta vector with the shifted version 
$(\delta_i,\delta_{i+1},\delta_i+\delta_{i+2},\delta_{i+1}+\delta_{i+3})$.
\item Shift the vector by one integer
$(\delta_i,\delta_{i+1},\delta_i+\delta_{i+2},\delta_{i+1}+\delta_{i+3}) \to  (0,\delta_i,\delta_{i+1},\delta_i+\delta_{i+2})$.
\item Add  the previous vector with the shifted version 
$(\delta_i,\delta_i+\delta_{i+1},\delta_i+\delta_{i+1}+\delta_{i+2},\delta_i+\delta_{i+1}+\delta_{i+2}+\delta_{i+3})$.
\end{enumerate}
This vectorized approach can be much more efficient than a naive scalar implementation~\cite{SPE:SPE2326}.

\paragraph{Core functions}
Beside compression and decompression, all our schemes need to support at least three core functions: 
\begin{itemize}
\item In a selection, we seek  the $i^{\mathrm{th}}$~integer. While we can always fully decompress all the integers and seek the result in the uncompressed array, 
we can typically  do much better with a specialized function that avoids unnecessary data manipulation---ideally accessing only the necessary data and avoiding committing to memory intermediate register values. 
\item We assume that the integers are stored in sorted order, and we want to seek the location of the first  value greater or equal to a given target, and to retrieve this value. Again, though we can implement this function by fully decompressing the data, we can  often do much better by avoiding a full decompression.
\item We also need to implement an in-order insertion. That is, under the assumption that the values are in sorted order, we need to be able to add one more value, while maintaining the result in sorted order. 
\end{itemize}
In all three cases, we are targeting relatively small compressed blocks (e.g., 256~integers). In instances where there are many more integers, a search might rely on auxiliary data structures. 
In our case, the B+-tree itself provides the 
indexing so that all the compressed data can be assumed to fit CPU cache.

\paragraph{Delete stability} \label{sec:deletestabl} All our schemes, except for binary packing  (BP128), satisfy a property that we call \emph{delete stability}: the removal of a value may not increase the storage requirement. It  was called the ``Delete Safe Property'' by Bhattacharjee et al.~\cite{db2luw2009} in the context of IBM~DB2 where it 
is  required as a design principle.
 Without this property, we may get the possibly unexpected effect that removing a key increases the storage requirement. To see why delete stability should not be taken from granted, consider the list of integers $\{1,2,3,4,\ldots \}$. Their successive differences are $\{1,1,1,1,\ldots \}$. However, suppose that we delete the second value, getting $\{1,3,4,\ldots\}$, then 
the successive differences contain a larger value (2): $\{1,2,1,1,\ldots \}$. Because the differences contain a larger value, they might be less compressible---depending on the compression algorithm used.

\subsection{VByte}

VByte~\cite{1559877,buttcher2010information,Stepanov:2011:SDP:2063576.2063627,Transier:2010:EBA:1877766.1877768,williams1999compressing} is one of the most popular and established integer compression techniques.
It is also known as variable-byte, var-byte, varint and escaping. 
We find it in common interchange formats (such as Google's Protocol Buffer) as well
as in search engines (such as Apache Lucene).
Starting from the least significant bits, we write non-negative integers using  seven bits in each byte, 
with the most significant bit of each byte set to 0 (for the last byte of an integer), or to 1. Integers in  $[0,2^7)$ are coded using a single byte, integers in $[2^7, 2^{14})$ are coded using two bytes and so on. See Table~\ref{table:illustration} for examples. There are many inconsequential variations on this format, e.g., we could use the value~1 to indicate the last byte of an integer instead of the value~0.

  \begin{table}
  \caption{\label{table:illustration}VByte compression of various integer values. The most significant bit of each byte is in bold. }
  \centering
  \begin{tabular}{cl} \toprule
  integer & VByte \\ \midrule
  1  &  \textbf{0}0000001\\
2  & \textbf{0}0000010\\
128 & \textbf{1}0000000,  \textbf{0}0000001\\ 
256  & \textbf{1}0000000,  \textbf{0}0000010 \\
32768 & \textbf{1}0000000,  \textbf{1}0000000,  \textbf{0}0000010\\
\bottomrule
  \end{tabular}
  \end{table}

For large 32-bit input values ($>2^{28}$), VByte is inefficient because it
requires more storage (5~bytes) than the uncompressed value (4~bytes). However, 
such a problem is uncommon if we use differential coding on sorted inputs.

Our VByte implementation uses standard C code.
VByte decoding can be fast when most integers fit in a single byte. One can then
decode over a billion integers per second on commodity superscalar desktop processors. 
However the performance can be lower when integers fit in various numbers of bytes, as
the cost of branch mispredictions increases. Indeed, the decode function requires
 frequent CPU branches when checking the continuation bits and there are data dependencies
when decoding a value.   In the worst case, the CPU
needs to process bytes sequentially.
Several other schemes were created to overcome this limitation, two of them were included in our tests: Masked~VByte~\cite{maskedvbyte} and VarIntGB~\cite{DeanOfficialplusslides:2009:CBL:1498759.1498761}.

Implementing a fast select or fast (sequential) search over VByte---without first decompressing all integers---is relatively straightforward. Moreover,
VByte can support fast insertions. That is, suppose that we want to insert value $a$ that would first between values $x_i$ and $x_{i+1}$. There is no need to change any of the bytes corresponding to the values $x_1, \ldots, x_i$. After these bytes, the data corresponding to $x_{i+1}-x_i$ must be updated so that we can store the bytes corresponding to $a-x_i$ and $x_{i+1}-a$.
Meanwhile, all the bytes corresponding to the values $x_{i+2}, x_{i+3},\ldots$ do not need to be modified, so that they can be merely moved in memory. This observation is not novel: 
B\"uttcher and Clarke remark that we can often merge two variable-byte stream without having to recompress them~\cite{buttcher}.
This makes VByte convenient from an engineering point of view.

To summarize, VByte's strength is its simplicity and convenience. It can be implemented
in just a few lines of code. Copying and splitting encoded sequences do
not require a re-encoding of the data which enables fast operations on
compressed values, i.e., to insert or delete values.

\subsection{VarIntGB}

VarIntGB~\cite{DeanOfficialplusslides:2009:CBL:1498759.1498761}
was engineered by Google to alleviate the performance problems that VByte
was causing. The main insight is that instead of encoding and decoding integers 
one at a time, we can encode and decode blocks of 4~integers instead. 

In VarIntGB, we use $\lceil \log_{2^8} (x + 1) \rceil $~continuous bytes to 
store an integer $x$. That is, if $x$ is in $[0,2^8)$, we use one~byte,
if $x$ is in $[2^8,2^{16})$, we use two~bytes, and so forth. 
Given four integers, $x_1, x_2, x_3, x_4$, we first compute 
the array of byte widths $M=\lceil \log_{2^8} (x_1 + 1) \rceil, \lceil \log_{2^8} (x_2 + 1) \rceil,
\lceil \log_{2^8} (x_3 + 1) \rceil,
\lceil \log_{2^8} (x_4 + 1) \rceil$. This array is made of four~integers in
$\{1,2,3,4\}$. We can store each of these values using 2~bits. Thus, the array
$M$ can be stored in a single byte (8~bits). The VarIntGB format encodes
each block of four integers using one byte for the array $M$, and then $M_1+M_2+M_3+M_3$~bytes
to represent the data corresponding to integers $x_1, x_2, x_3, x_4$ respectively. See Fig.~\ref{fig:groupvarint}. In practice, if the number of integers is not divisible by four, we may end with a partial block.

When decoding VarIntGB data, we first examine the byte $M$, and extract its four components. The integers $x_1, x_2, x_3, x_4$ are then decoded from the following $M_1+M_2+M_3+M_3$~bytes. Unlike VByte,  the decoding of VarIntGB requires a fixed number of operations per encoded integer  in our implementation. Thus, VarIntGB can be expected to offer a better performance than VByte over poorly compressible data. Moreover, we can address the important case where all four integers are stored using a single byte, and decode the four integers quickly, thus achieving a superior performance for highly compressible data.

\begin{figure}
\caption{\label{fig:groupvarint}Encoded bytes corresponding to integer values $1024,12,10,512$ using VarIntGB\@. The result occupies seven bytes.}
\centering\begin{tikzpicture}[node distance=0cm,start chain=1 going right]   \tikzstyle{mytape}=[draw,minimum height=1cm]
    \node  [on chain=1,mytape,fill=red!20] {01|00|00|01};
    \node  [on chain=1,mytape,fill=green!20] {0x0004};
    \node  [on chain=1,mytape,fill=green!20] {0xc};
    \node  [on chain=1,mytape,fill=green!20] {0xa};
    \node  [on chain=1,mytape,fill=green!20] {0x0200};
\end{tikzpicture}
\centering
\end{figure}

Like VByte, implementing a fast select or a fast sequential search over VarIntGB is straightforward. However,
a fast insertion is more difficult. If we find that the values must be inserted at index $i$, then we can certainly avoid rewriting the first $\lceil i/4 \rceil$~values. However, recoding 
remaining values in-place following the insertion of  a single new value is both technically cumbersome and relatively slow. We found it more appropriate to  decompress the remaining values, and then recompress them with the new value added. 

\subsection{Masked~VByte}

Our VByte decoder algorithmically processes one input byte at a time.
Though VarIntGB accelerates the processing by changing the format,
it might also be possible to accelerate VByte itself by designing a SIMD-based
decoder. Indeed, the VByte format might be otherwise convenient, if only
because  it is a well-known and time-tested format.
 Stepanov et al.~\cite{Stepanov:2011:SDP:2063576.2063627} tested such a
 SIMD-based decoder but got only modest gains (less than \SI{25}{\percent}).
 However, Plaisance et al.~\cite{maskedvbyte} showed for the first time that
 it is possible to multiply VByte decoding speed by using SIMD instructions
 in a decoder called Masked~VByte.
 
The Masked~VByte approach first gathers the most significant bits  of an array of consecutive bytes. A single instruction (\texttt{pmovmskb}) can serve for this purpose on Intel processors. To illustrate the approach, suppose we code the integers 128, 386, 16, 32 using VByte. The result will be six compressed bytes: {\textbf{1}0000000}, {\textbf{0}0000001}, {\textbf{1}0000010}, {\textbf{0}0000011}, {\textbf{0}0010000}, and {\textbf{0}0100000}. We can gather the control bits (1,0,1,0,0,0 in this case) using the \texttt{pmovmskb} instruction.
We can then use another instruction (\texttt{pshufb}) that permutes the bytes of a register in a desired way (that may
be specified using a control mask) in as little as one cycle. Finally, we need to mask or shift the  bytes to arrive at the decoded integers. 
The exact algorithm and its implementation
is technical and requires many steps, so we refer the reader to the original work
for details~\cite{maskedvbyte}.

The Masked~VByte can be three times faster than a conventional VByte decoder when the data is 
sufficiently compressible. Masked~VByte also benefits from an integrated SIMD-accelerated differential coding.

We implemented  accelerated select and sequential search functions that are similar to the VByte functions. 
Selection is SIMD-accelerated: we decode in registers the first $4 \lceil i/4 \rceil $~values and return the value at the proper index. Sequential search is handled similarly. For the insertion, we use the same function as for VByte: this is possible because the underlying data format is identical.

\subsection{Binary Packing (BP128)}

The binary representation of an unsigned integer $x$
has all but the less significant $\lceil \log_2 (x+1)\rceil$~bits 
equal to zero.
Given an array of integers $x_1, x_2, \ldots$, if $m$ is the maximum value ($m=\max x_1, x_2, \ldots$), then we can store all integers using $\lceil \log_2 (m+1)\rceil$~bits per integer. Effectively, we can truncate the leading zeros.
Binary packing is a compression technique that exploits this idea. We regroup
all integers in blocks of, say, 128~integers. We find $b=\lceil \log_2 (m+1)\rceil$ and store it using as little as a byte, and then we write out the 128~integers using $b$~bits per integer, packing them tightly so that $128b$~bits are used in total.

Like with VarIntGB, if the number of integers is not divisible by 128~integers, we may end with a partial block. For simplicity, we may pad the input with zeros so that the number of integers is  divisible by 128.

Lemire et al.~\cite{SPE:SPE2326} describe an optimized SIMD-accelerated version of binary packing, henceforth BP128, where blocks of 128~integers are used. Differential coding is integrated in the unpacking routines. 

We implemented fast functions to select just a single value, that is, if the $i^{\mathrm{th}}$ value is sought, only the first $4 \lceil i/4 \rceil$~values are decoded (in registers) and the desired value is returned. 

We implemented a similar sequential search function. Insertions require decoding the entire array, editing the uncompressed data and recompressing. 

\subsection{Frame of Reference}

Though all compression techniques presented so far rely on differential 
coding, there are other alternatives. Indeed, differential coding has the downside that it is relatively difficult to random skip values: without auxiliary data structures, one must decode integers starting from the beginning. In particular, it might be difficult to use a fast
search technique, like binary search, directly on the compressed data.

One convenient compression  alternative that offers good compression as well as fast random access is 
 Frame-Of-Reference (FOR)~\cite{655800}. In FOR, arrays of values are partitioned into blocks (e.g., of 32~integers). 
We code the minimal value of the block, and then all values
are written in reference to this minimum. For example, the block of values $\{500, 521, 531, 574\}$ would be written as $\{21, 31, 74\}$. To decode these values, all we need is the minimum (500), and then we can compute the sums $\{500, 21+500, 31+500, 74+500\}$.
When the arrays are sorted, as  in our application, the minimal value is always the first one.  
Thus, starting from the array of sorted values $x_1, x_2, x_3,\ldots, x_n$, we pack the array $x_1-x_1,x_2-x_1, x_3-x_1,\ldots, x_n- x_1$ of transformed integers using $\lceil\log_2 (x_n-x_1 +1)\rceil$~bits per integer, packing and unpacking them quickly, as in binary packing. 
 In our implementation, we use blocks that are multiples of 32~integers for the scalar version (henceforth FOR) and multiple of 128~integers for the SIMD-accelerated version (henceforth SIMD~FOR). When the number of input integers is not divisible by the block size, we create a partial block, packing only the necessary $x$~integers, to maximize the compression ratio.

One can select the $i^{\mathrm{th}}$~value in a block in constant time (irrespective of the block size). If the values are sorted, searching for a value can be done in logarithmic time using binary search. Inserting a new value can be done by uncompressing the block, inserting the value in the uncompressed data and then recompressing it.

\section{Database Compression in Upscaledb}

Upscaledb is an embedded key-value database engine implemented in C++. It is
used for a variety of tasks like caching web crawler data, gathering and
pre-processing sensor input and network events, storing the metadata of
backup software or just as a general data store for mobile and desktop
applications.

Upscaledb's functionality is similar to other key-value database engines like
Berkeley~DB, but different in its implementation. Instead of using type-less
byte arrays as keys, Upscaledb is aware of the key type. It can therefore
optimize the underlying data structures and algorithms for the type.

The key type is specified by the user when creating a database. Among the
supported types are 32-bit integers, 64-bit integers, variable-length binary
data and fixed-length binary data. This configuration  parametrizes
C++ template classes for lower-level structures and algorithms.

\begin{figure}[h]
\centering\includegraphics[width=0.99\columnwidth]{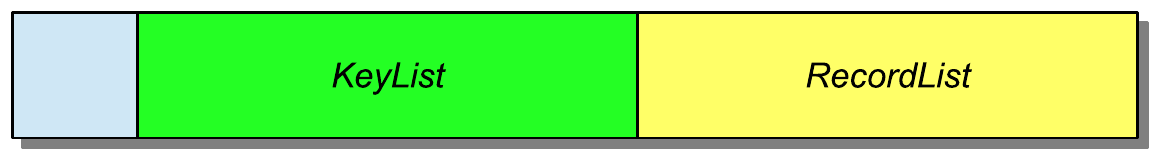}
\caption{\label{fig:layout}Memory layout of a B+-tree node}
\end{figure}

\begin{figure}[h]
\centering\includegraphics[width=0.99\columnwidth]{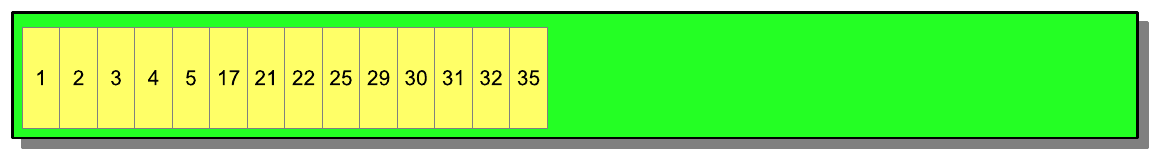}
\caption{\label{fig:layout-fixlen}The \texttt{KeyList} storing a sorted, non-consecutive sequence of fixed-length integer keys}
\end{figure}

\begin{figure}[h]
\centering\includegraphics[width=0.99\columnwidth]{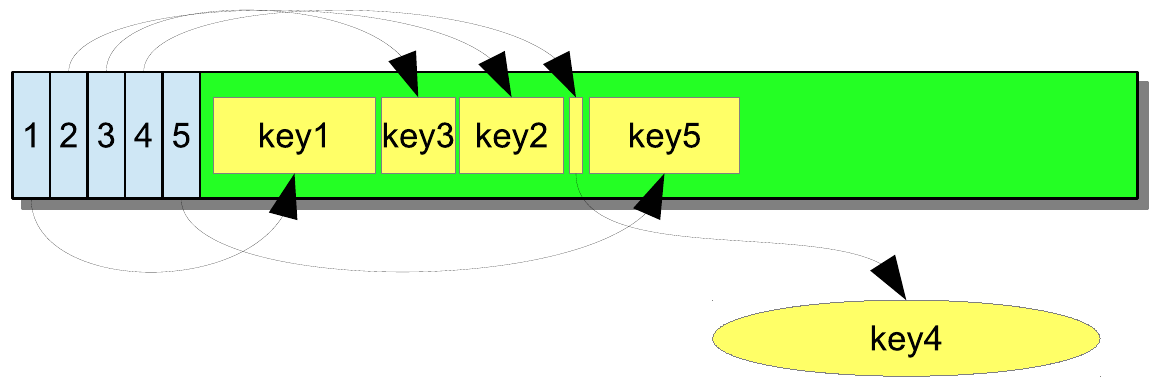}
\caption{\label{fig:layout-varlen}The \texttt{KeyList} storing variable-length keys}
\end{figure}

Upscaledb uses a B+-tree~\cite{Comer:1979:UB:356770.356776} structure to store
the database indices. A B+-tree is similar to a conventional B-tree.
It is an associative map that stores keys and values in sorted order and supports
logarithmic-time search, insertion and deletion operations.
However, the B+-tree 
stores values aligned with the leaf nodes. Moreover, the leaf
nodes form a linked list for fast traversal.

B+-trees are a common data structure.
Indeed, B+-trees are also used by other key-value stores like Berkeley~DB and Tokyo Cabinet, as well as many other database engines such as Oracle and SAP~HANA.

Upscaledb's B+-tree node (also called a \emph{page}) stores keys and values separately from each other.
The actual in-memory layout is described in Fig.~\ref{fig:layout}. Each node has a header
structure of 32~bytes containing flags, a key counter, pointers to the
left and right siblings and to the child node. This header is followed by the
\texttt{KeyList} (where we store the key data) and the \texttt{RecordList} (where we store the value's data).
Their layout depends on the index configuration and 
data types.

The \texttt{RecordList} of an internal node stores 64-bit pointers to child nodes,
whereas the \texttt{RecordList} of a leaf
node stores values or 64-bit pointers to external blobs if the values
are too large.

 Fixed-length keys (Fig.~\ref{fig:layout-fixlen}) are always stored sequentially and without
overhead. They are implemented as plain C arrays of the key type, i.e.,
\texttt{uint32\_t keys[]} for a database of 32-bit integers. Variable-length
keys (Fig.~\ref{fig:layout-varlen}) use a small in-node index to manage the keys. Long keys are stored
in separate blobs; the B+-tree node then points to this blob.

The Upscaledb in-memory representation of fixed-length keys eases the application of compression.
The keys are already stored in sorted order, therefore applying
differential encoding does not require a change in the memory layout.
Since keys are stored sequentially in memory, SIMD instructions can be used
efficiently if needed.

\subsection{Insertion and Deletion}

In a conventional B+-tree, non-root, non-leaf nodes can accommodate between $b$ and $2b$~keys.  Leaf nodes accommodate between
$b$ and $2b-1$~keys. Nodes are split or merged following insertions and deletions to remain
in these ranges. (See~\ref{appendix:insertionanddeletions} for details.)

Upscaledb differs from a conventional B+-tree in two significant ways.
Firstly, the capacity of nodes is defined in terms of storage space,
and not as a number of keys. Secondly, the nodes are balanced locally:
the result of an insertion or deletion does not immediately 
propagate back to
the root of the tree. We review these two differences in more details in
the rest of this section.

\begin{description}
\item [Capacity as storage space] 
Instead of defining capacity as the number of keys that a node
may include, Upscaledb fixes the maximal space usage of a node
(16\,kB by default). This space is used to store both keys and
values. 
Unlike conventional B+-trees that forbid  nodes from being less than half
full, Upscaledb only considers merging nodes that are nearly empty, i.e.,
nodes that have less than 4~keys.

Leaf nodes use compression, and the maximal number of keys that
can be stored depends  on the compressibility of the data. The size
of the value entries has also an impact, since they both share
the same space.

When inserting a new key in a node with compressed keys, it might be necessary to
attempt the insertion to determine how many bytes the new node would use.
Only then might we determine that the node is full and needs to be split.

Upscaledb follows this approach. If a key cannot be stored in the current
space allocated to the \texttt{KeyList} then it tries to reorganize the node (e.g., 
by growing the \texttt{KeyList} at the expense of the \texttt{RecordList}, if possible).
As a last resort, the node is split and the
insert operation is done in one of the new nodes.

As mentioned in \S~\ref{sec:codecs}, the binary packing schemes violate the principle of
``delete stability''. A BP128-encoded block can grow if an integer is deleted
from the block. This can lead to cases where deleting a key from
a B+-tree leaf causes an overflow and triggers a node split. 
To our knowledge, Upscaledb is unique among B+-tree implementations in that it can split nodes when deleting keys. Such an ability allows us to use BP128 as a codec.

In a B+-tree, most nodes are leaf nodes: there would be little
storage gain in compressing non-leaf nodes.
For example, to store 20~million keys, we might use 4~non-leaf nodes and 2500~leaf nodes.
Consequently, 
in Upscaledb, only leaf nodes use compression. Thus,
deleting a key from a non-leaf node cannot lead to this node being split. IBM~DB2 also compresses only leaf nodes~\cite{db2luw2009}, for the same reason.

\item [Local balancing]

When inserting a new key, conventional B+-tree implementations usually first go down the tree to find the right node, and then potentially split the nodes from the bottom up. Moreover, when deleting a key, underfilled nodes are balanced. Splits and balancing operations can require modifications in the B+-tree which are propagated all the way up to the root level.

Upscaledb's implementation is different as it is balancing is only performed locally. It follows 
Guibas and Sedgewick~\cite{Guibas:1978:DFB:1382432.1382565} and examines nodes while descending the tree from the root.
Nodes that cannot accommodate a new key are split and 
underfilled nodes are merged---without propagating changes above the parent.
This local balancing may improve query latency.

This localized approach works because any parent node is guaranteed to have
space for one more key---irrespective of the value of the new key. Hence, when we split the current node, there is no need to immediately split the parent to make room for one more key. There are two  minor inconveniences to this localized approach: it can happen that nodes are split although the split is not immediately required. The other inconvenience is that empty nodes are not pruned if their removal would cause global updates in the B+-tree, since we want to restrict our updates to a local scope only. However, this waste only happens in rare situations, and these empty nodes represent only a small fraction of the total storage.

When descending the B+-tree toward the leaf node, the insert and delete algorithms 
are identical in their implementation.  Only when the leaf is reached, do the algorithms
diverge and perform their specific action to either insert or delete a key.

If the operation aborts because a node split is required
then the node is split, the parent node and the siblings are updated---without propagating the changes further. 
This is the case even for deletion operations, where
some compression codecs might actually require more space when keys are deleted.

\end{description}

\subsection{Integrating Compression}

Upscaledb implements the typical operations of key-value databases like
inserting or overwriting keys, deleting keys and looking up keys.
In addition, bi-directional cursors can iterate over the keys. When there is no compression, many of these operations follow the same pattern: binary search is
used to find the position of the (new) key in the node; then the operation is
executed at the specified position. This works well because we have very fast random access over uncompressed arrays of keys. With integer compression, more care is needed.

To improve performance, the compressed integers are split in blocks.
Therefore, we have the following hierarchy: each leaf node contains a \texttt{KeyList}
which may contain several blocks. The block size depends on the codec. 
BP128 stores up to 128~integers per block, all other codecs store up
to 256~integers by default. We arrived at these block sizes through empirical evaluation (see \S~\ref{sec:blocksize}).

Each block is described by a small index structure at the beginning of the \texttt{KeyList},
containing the offset of the block in bytes relative to the beginning of the \texttt{KeyList}, the number of keys in the block,
and the size of the block in bytes (or, in case of BP128, the number of
bits required for encoding). 

Also, each block stores the start value of the
encoded integers. This value serves as the starting value when decoding 
the differentially coded values in the block
and it is used to locate the block of a given key. Blocks can grow, but as soon
as they reach a limit they are split. 

Blocks are stored sequentially
within the \texttt{KeyList}. Following insertions and deletions, blocks
can contain various numbers of keys. It is even possible for a block to become
empty (to become a \emph{gap}).  However, the space used by a block is not
necessarily reclaimed eagerly as its content is reduced.

If the B+-tree node overflows, it is split. Since this is an expensive
operation, several attempts are made to optimize the node's layout and save
space,  to delay the actual split. Blocks are reorganized and gaps
are removed. Also, the space which is assigned to the \texttt{RecordList} can be reduced,
and assigned to the \texttt{KeyList}, and vice versa.

From the following list, each integer codec has to support the
\texttt{compress} and \texttt{decompress} functions. The other functions are
not mandatory.

\begin{itemize}
\item Compress --- Compresses a block of integers to a provided memory location.
\begin{lstlisting}
uint32_t compress_block(Index *index, const uint32_t *in, uint32_t *out);
\end{lstlisting}
The \texttt{compress} function returns the number of bytes required to
compress the data.
\item Decompress --- Decompresses a block of compressed integers to a buffer.
\begin{lstlisting}
void decompress_block(Index *index, const uint32_t *in, uint32_t *out);
\end{lstlisting}
\item Insert --- Inserts a new key in a compressed block. Except for
        BP128, FOR and SIMD~FOR, all codecs provide such a custom insert
        function: see \S~\ref{sec:fastupdates} for details.  BP128, FOR and SIMD~FOR decode the block, modify the
        decompressed data and re-encode the block.
\begin{lstlisting}
bool insert(Index *index, uint32_t *in, uint32_t key, uint32_t *pslot);
\end{lstlisting}
The insert functions returns false if the key already exists. It also returns
the position of the new key in the \texttt{KeyList}. This position (the ``slot'') is then
required to insert the corresponding value into the \texttt{RecordList}. It is assumed
that the block has enough free space to insert another key. Growing or splitting
the block is handled by the caller.

If possible, the insert function  implements a fast code path to append
a new key at the end. See \S~\ref{sec:fastappend} for details.

\item Find --- Performs a lower bound find for a key, returns the location of the first value at least as large as the specified key value. This function is used to search for a key.
\begin{lstlisting}
int find_lower_bound(Index *index, const uint32_t *in, uint32_t key, uint32_t *presult);
\end{lstlisting}

\item Delete --- Deletes a key. Only implemented by the VByte and Masked~VByte codecs. Other
        codecs decode the block, modify the decompressed data and re-encode
        the block. 
\begin{lstlisting}
template<typename GrowHandler>
void del(Index *index, uint32_t *in, int slot,
            GrowHandler *handler);
\end{lstlisting}
The codecs do not necessarily have ``delete stability'' (see \S~\ref{sec:deletestabl})
and can require more space after a key is deleted. The \texttt{GrowHandler} template
parameter is used to signal such circumstances to the caller. It can assign
more space to the current block or request a B+-tree node split.

\item Vacuumize --- Reorganizes all blocks, trying to reduce gaps and space
        to avoid B+-tree node splits. 
\begin{lstlisting}
void vacuumize();
\end{lstlisting}

\begin{figure}
\centering\includegraphics[width=0.99\columnwidth]{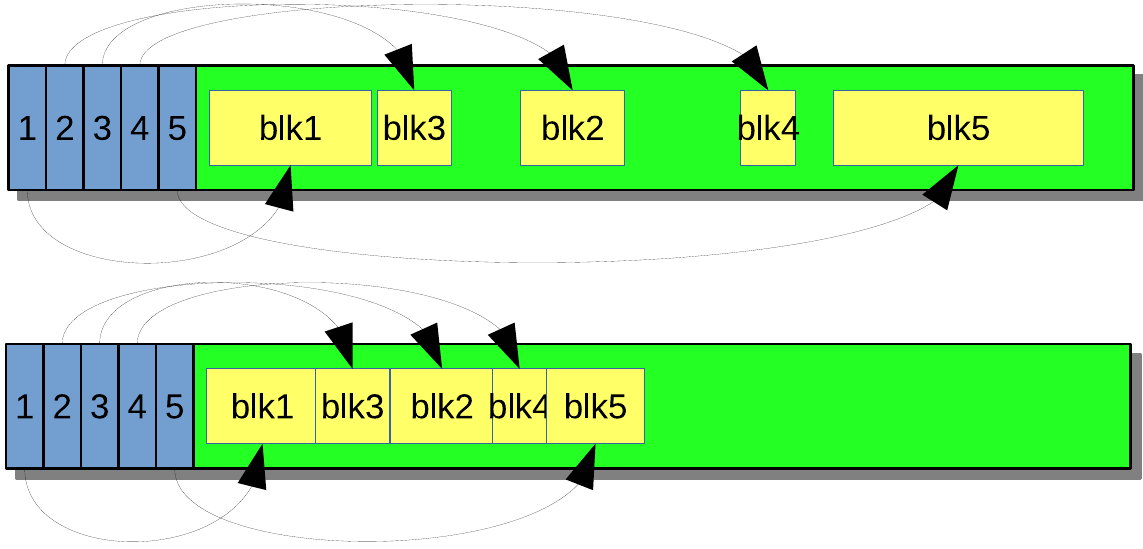}
\caption{\label{fig:vacuumize}A \texttt{KeyList} before (top) and after (bottom) the
vacuumize-operation. Fragmentation occurs when empty blocks are deleted. Blocks will have unused space if keys are deleted from a block.
}
\end{figure}

When ``vacuumizing'' a \texttt{KeyList} (see Fig.~\ref{fig:vacuumize}), the BP128,
FOR and SIMD~FOR codecs decode all blocks into temporary
memory and re-encodes them into new (usually fewer, densely packed) blocks.
The other codecs just move the blocks to remove any gaps between the blocks. 
\end{itemize}

\subsection{Fast In-place Updates}

\label{sec:fastupdates}

We can  shift an array of bytes in memory by a byte offset at high speed. The C~language offers the \texttt{memmove} function for this purpose, and it
is highly optimized---to the point of being limited by the memory throughput.
Thus, it is efficient to modify byte-oriented formats in-place (e.g., VByte and VarIntGB).
However, we are not aware of any similarly fast function (within a factor of four) to shift
an array of bytes by a bit offset that is not divisible by eight. Thus, inserting a value that occupies a number of bits non-divisible by eight in a packed array 
is likely to be a relatively slow operation. This makes BP128, FOR and SIMD~FOR data streams more difficult to update in-place.

In our implementation, VByte, VarIntGB, and Masked~VByte  perform all update operations directly on compressed data. 
BP128, FOR and SIMD~FOR still
require a decode-modify-encode loop for updates.

\subsection{Fast Append Functions}
\label{sec:fastappend}

A common database index operation is to insert new keys at the end, i.e.,
for time-series data where the key is a chronologically incremented timestamp.
Appending integers to an array of differentially-coded compressed integers could be slow if we had to first decode the last value (by summing up the differences).
To improve performance, the block descriptor stores the value of the last integer. With this cached value, it is trivial to
decide whether a new key is appended at the end or inserted in the middle. If
it is appended, then its delta value can be calculated by subtracting the
(previously) highest block value from the new value. This optimization
improved performance by up to about \SI{30}{\percent} for the insertion of sequentially ordered keys.
Adding 32~extra bits to the descriptor does not degrade the compression ratios by a significant degree when dozens or even hundreds of keys are stored in a block.

All codecs support fast append functions that avoid
first uncompressing the data at least some of the time. For BP128, we only uncompress
the block if the existing bit width is insufficient compared to the size
of the new delta to be appended. Otherwise, we modify the compressed
data directly. We proceed with FOR and SIMD~FOR similarly: if the existing bit width
is sufficient, we append directly in the compressed data, otherwise we are 
forced to first uncompress the block.

\section{Benchmarks}

We first benchmark separately the various operations
that are relevant to a key-value store: decompression
speed, insertions, select and search. After benchmarking
the operations separately, we then report
on the performance in Upscaledb with realistic data.

\subsection{Hardware}

All our experiments are executed on an Intel Core
i7-4770 CPU (Haswell) with 32\,GB memory (DDR3-1600 with double-channel). The CPU has 4 cores of 3.40\,GHz
each, and 8\,MB of L3 cache. Turbo Boost is disabled on the test machine, and the processor is set to run at its highest clock speed.
The computer runs Linux Ubuntu~14.04. We report wall-clock time.

Except for the
fact that dirty B+-tree nodes in Upscaledb are purged in the background, all
tests are single-threaded. Our workloads are small enough to
fit in memory and input-output is not a limiting factor.

\subsection{Microbenchmarks and Evaluation}

\label{sec:micro}

We compare the codecs---without database interaction.
 We
compile our C++ benchmarking software using the GNU GCC~4.8.2 compiler with  the \texttt{-O3} flag. 
Our implementation is freely available 
under an open-source license.\footnote{https://github.com/lemire/SIMDCompressionAndIntersection}

\begin{figure}[tbh]
\centering
\subfloat[Bits per integer \label{fig:bitsperinteger}]{\centering\includegraphics[width=0.99\columnwidth]{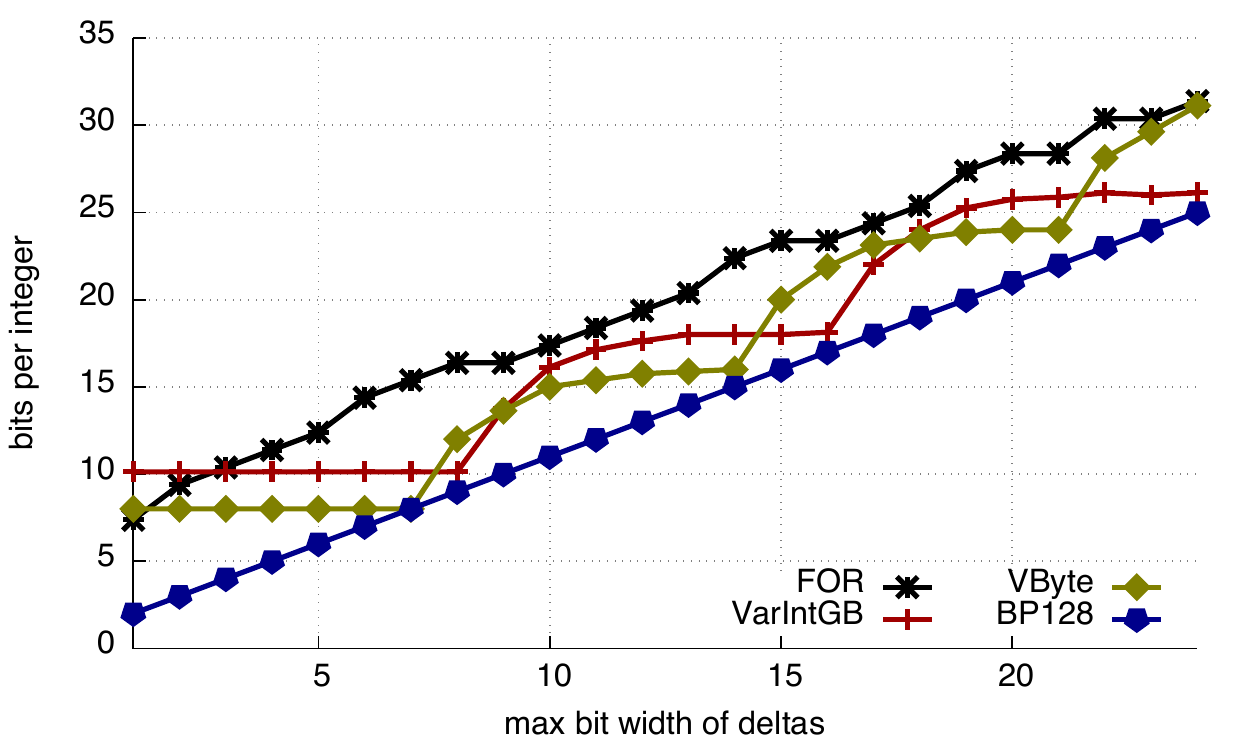}}
\\
\subfloat[Decompression speed (Bis) \label{fig:decomptime}]{\centering\includegraphics[width=0.99\columnwidth]{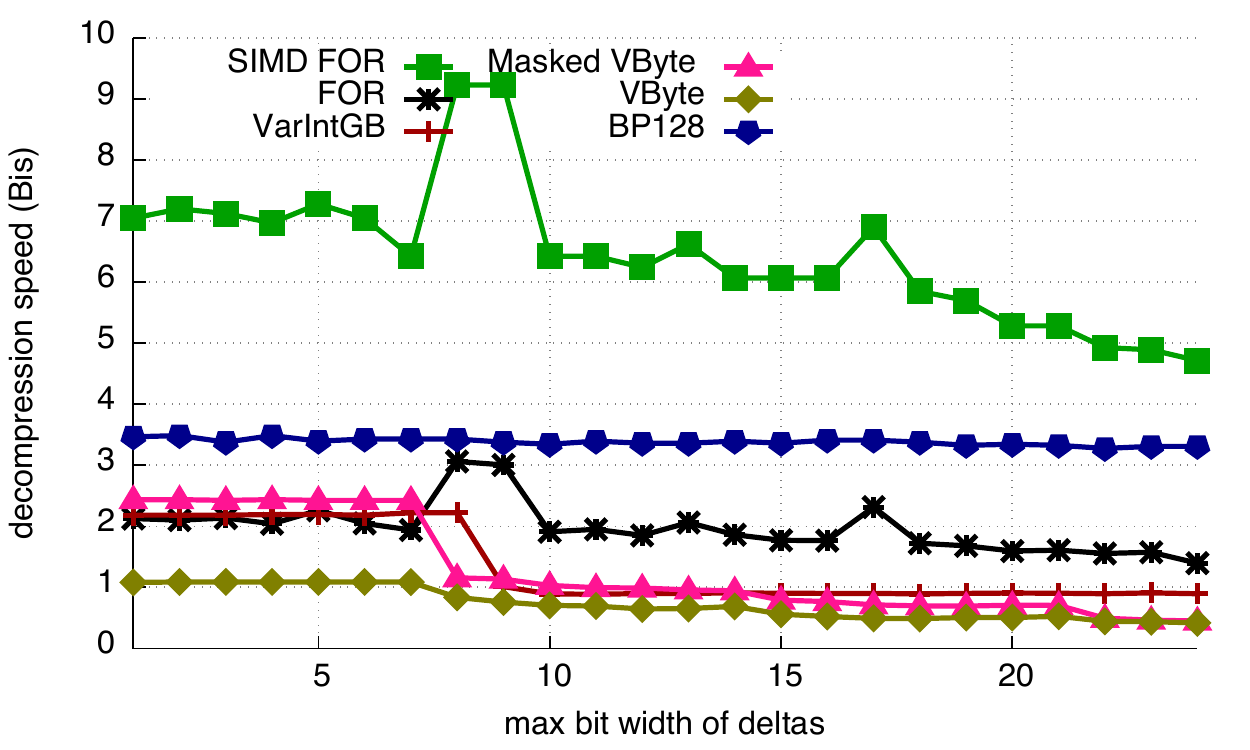}}
\caption{\label{fig:microcomp} Compression rates and decompression speed }
\end{figure}

Given a bit width $b \leq 24$, we first generate an array of 256~integers in $[0,2^b)$: $\delta_1, \delta_2, \ldots$. The prefix sum is computed ($\delta_1, \delta_1+ \delta_2, \ldots $) and used as input data. The result is a sorted list of 32-bit integers.
Fig.~\ref{fig:microcomp} shows the average compressed size (in bits per integer) and the decompression speed in billions of integers per second. We see that BP128 offers the best compression whereas FOR and SIMD~FOR offer poorer compression compared to other codecs. 

Regarding the decompression speed, we report the numbers  in billions of 32-bit integers decompressed per second (Bis). See in Fig.~\ref{fig:decomptime}. SIMD~FOR is twice as fast as the next scheme, BP128, which is itself much faster than most other alternatives (up to twice as fast). VByte is the only codec that is limited by a best speed of only about 1~billion integers per second. In contrast, SIMD~FOR can decompress data at a rate of over 7~billions integers per second---or about two integers decoded  per clock cycle.

\begin{figure*}[tbh]
\centering
\subfloat[Time to create 256-integer array from random inserts \label{fig:insert}]{\centering\includegraphics[width=0.45\textwidth]{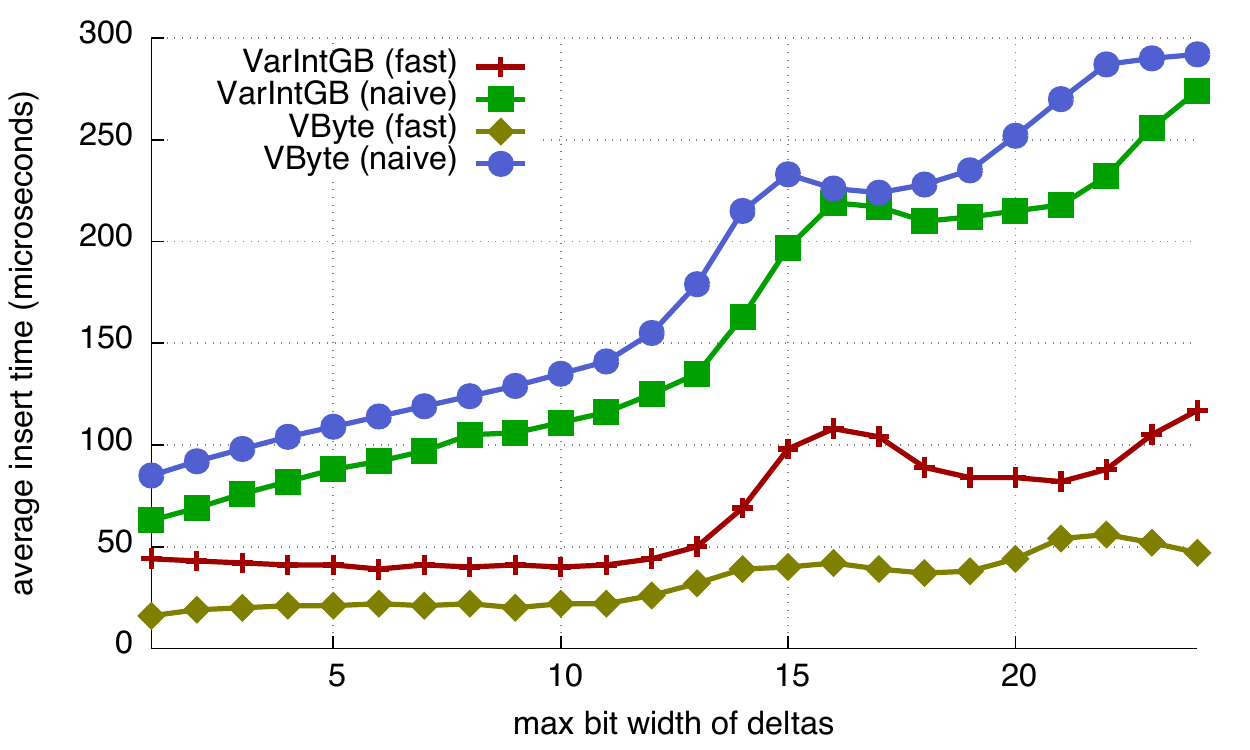}}
\subfloat[Select speed \label{fig:select}]{\centering\includegraphics[width=0.45\textwidth]{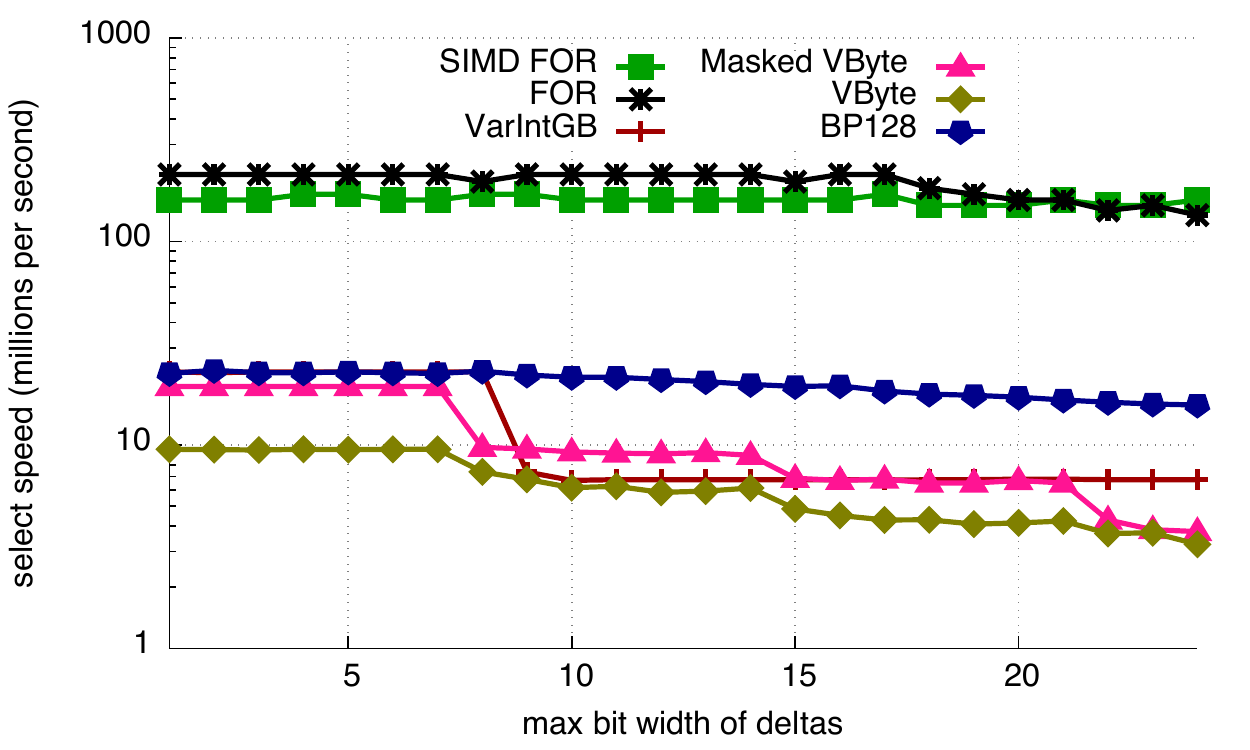}}

\subfloat[Find speed \label{fig:search}]{\centering\includegraphics[width=0.45\textwidth]{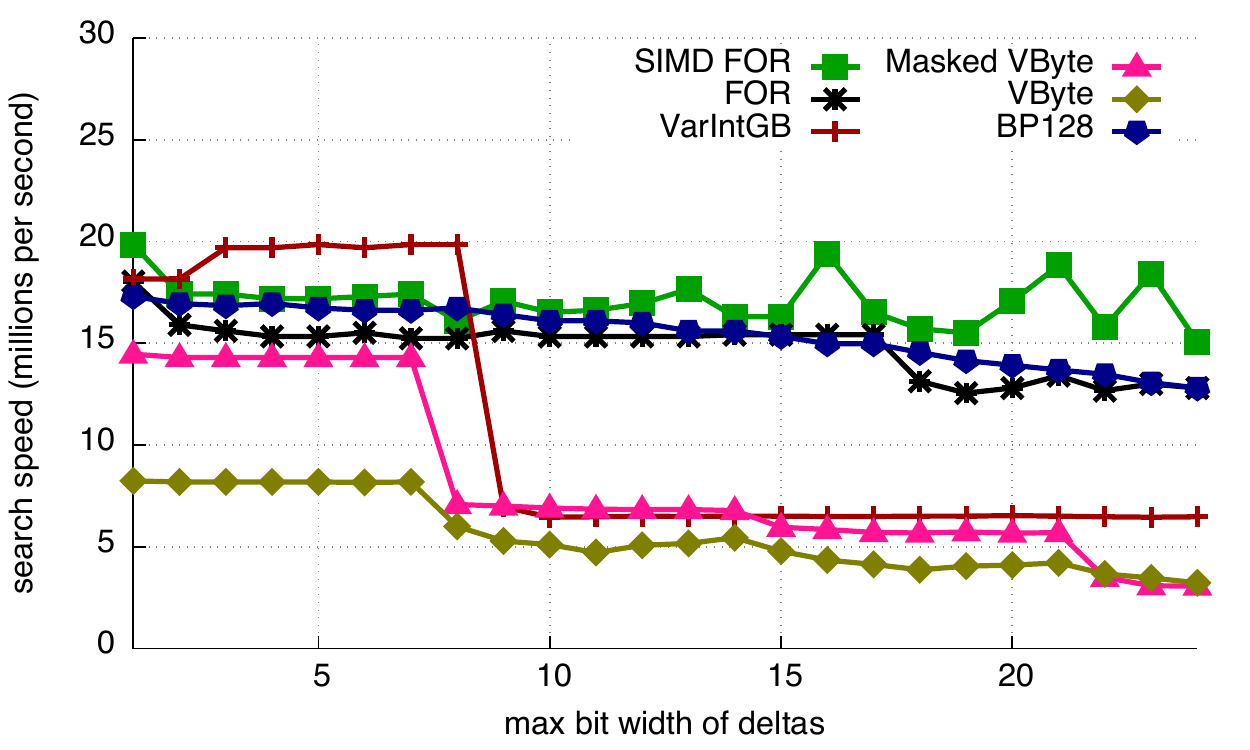}}
\caption{Operation timings and speeds over compressed data, for various codecs. \label{fig:micro} }
\end{figure*}

Fig.~\ref{fig:micro} presents various benchmarks regarding operations on compressed data.
\begin{itemize}
\item Instead of starting from the integers in sorted order and compressing them, we pick the integers at random one by one and we insert them in the VByte or VarIntGB stream (Fig.~\ref{fig:insert}). In the naive implementation, the stream is first decompressed, we insert the value and then recompress the stream. In the fast version, we use our optimized function. We see that the optimized function can be several times faster. We also see that VByte is slightly faster than VarIntGB  in this case due to its simpler data layout.
\item In Fig.~\ref{fig:select}, we randomly select the value at one of the indexes. We present the data in millions of operations per second with a logarithmic scale. We see that FOR and SIMD~FOR are an order of magnitude faster at this task because they do not rely on differential coding. BP128 is the next fastest codec while VByte is the slowest.
\item In Fig.~\ref{fig:search}, we benchmark the find function by randomly seeking a value in range. In this instance, all schemes but VByte are nearly as fast (within a factor of two) for compressible data, while VByte is significantly slower. VarIntGB offers the best performance in this case. FOR and SIMD-FOR differ from the other schemes in that they use a binary search (as a sequential search proved slower) whereas all other codecs rely on a sequential search. If the block size was much larger, FOR and SIMD~FOR could be expected to perform better, but we are not interested in that case.
\end{itemize}

Overall, our results suggest that on speed and compression ratios,  BP128 offers good performance. If faster random access is necessary, and the compression ratio is not an issue, then  FOR and SIMD-FOR might be preferable.

\subsection{In-database Benchmarks}

Though synthetic benchmarks show that some compression schemes
are superior than others, we are interested in the effect of compression 
in an actual key-value store.
For this purpose, we
compiled Upscaledb and our benchmarking software using the GNU GCC~4.8.2 compiler with  the \texttt{-O3} flag.  Upscaledb is freely available
under an open-source license (\url{http://upscaledb.com}).  The benchmarks are
executed with ``\texttt{ups\_bench}'', a benchmarking tool which is part of Upscaledb's
sources. 

We aim to study the compression of the integer keys. For this purpose, we only stored the keys, without any accompanying values. All our queries bear only on the keys.

For our experiments, we use the ClusterData model from Anh and Moffat~\cite{Anh:2010:ICU:1712666.1712668}. We vary the number of keys generated, up to
a billion. When generating $N$~keys, we set the range of possible values to $[0,9N/8)$.
We insert the keys in order. We choose this data distribution because it is a reasonable model for realistic data.

\begin{figure}
\centering\includegraphics[width=0.99\columnwidth]{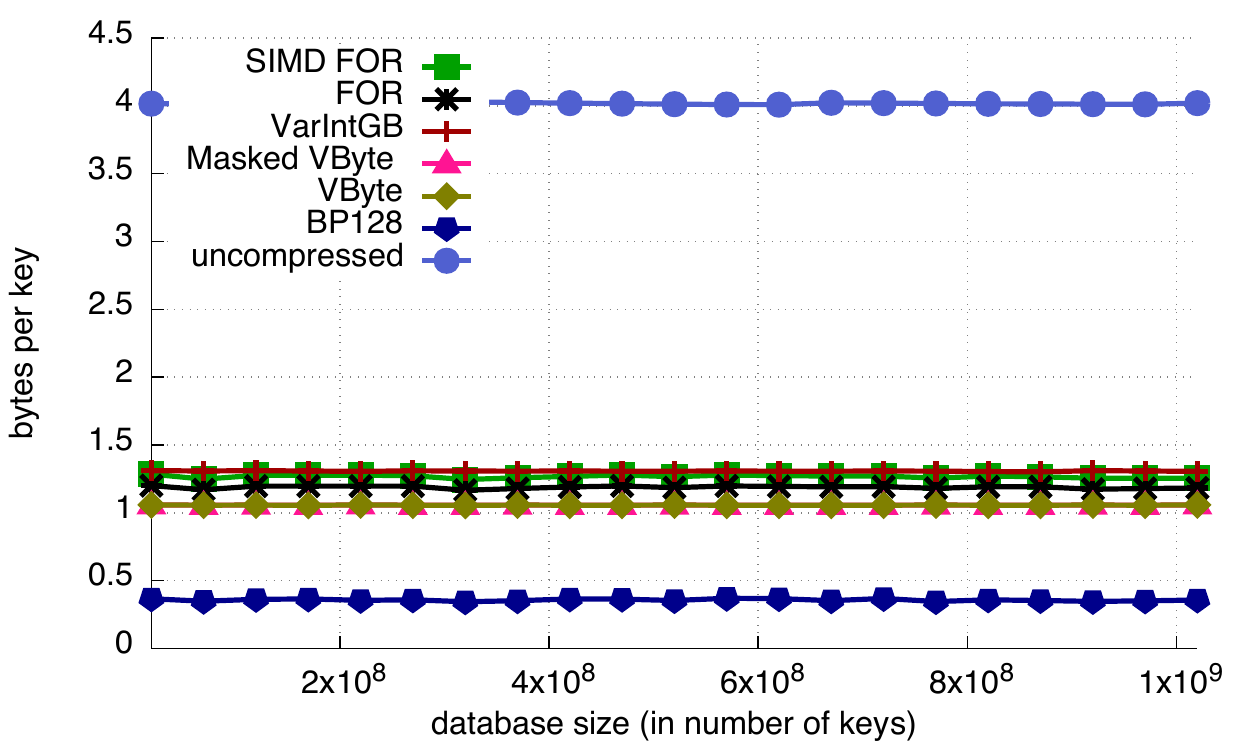}
\caption{Relative database sizes in Upscaledb (ClusterData)\label{fig:relsize}}
\end{figure}

Table~\ref{tab:dsize} and Fig.~\ref{fig:relsize} present compression results with various database sizes, using the default block sizes (128 for BP128 and 256 for other codecs, see \S~\ref{sec:blocksize}).
Expectedly, the best compression
is offered by BP128 which can compress the database by a factor of ten compared
to an uncompressed B+-tree. The compression ratios offered by the other codecs are similar (compression ratio of 2 or 3), with SIMD~FOR compressing slightly less and VByte compressing slightly better. Both VByte and Masked~VByte have exactly the same compressed output.
In these tests, we see that the compression as measured by the number of bytes used per key is nearly constant irrespective of the database size.

\begin{table}
\caption{\label{tab:dsize}Database sizes in bytes per key, for ClusterData model ($N=\num{20000000}$)}
\centering
\begin{tabular}{l|l}
\toprule
key format & bytes per key\\
\midrule
uncompressed & 4.02 \\
VByte/Masked~VByte & 1.06\\
VarIntGB &  1.31\\
FOR &  1.26\\
SIMD FOR &  1.28  \\
BP128 &  \textbf{0.37}\\
\bottomrule
\end{tabular}
\end{table}

\subsubsection{In-database timings}

Each benchmark runs three times, the median result is reported. The difference between the median and other timings is small (typically less than \SI{1}{\percent}). 

\begin{figure*}[tbh]
\centering
\subfloat[Relative lookups timings \label{fig:rellookup}]{\centering \includegraphics[width=0.99\columnwidth]{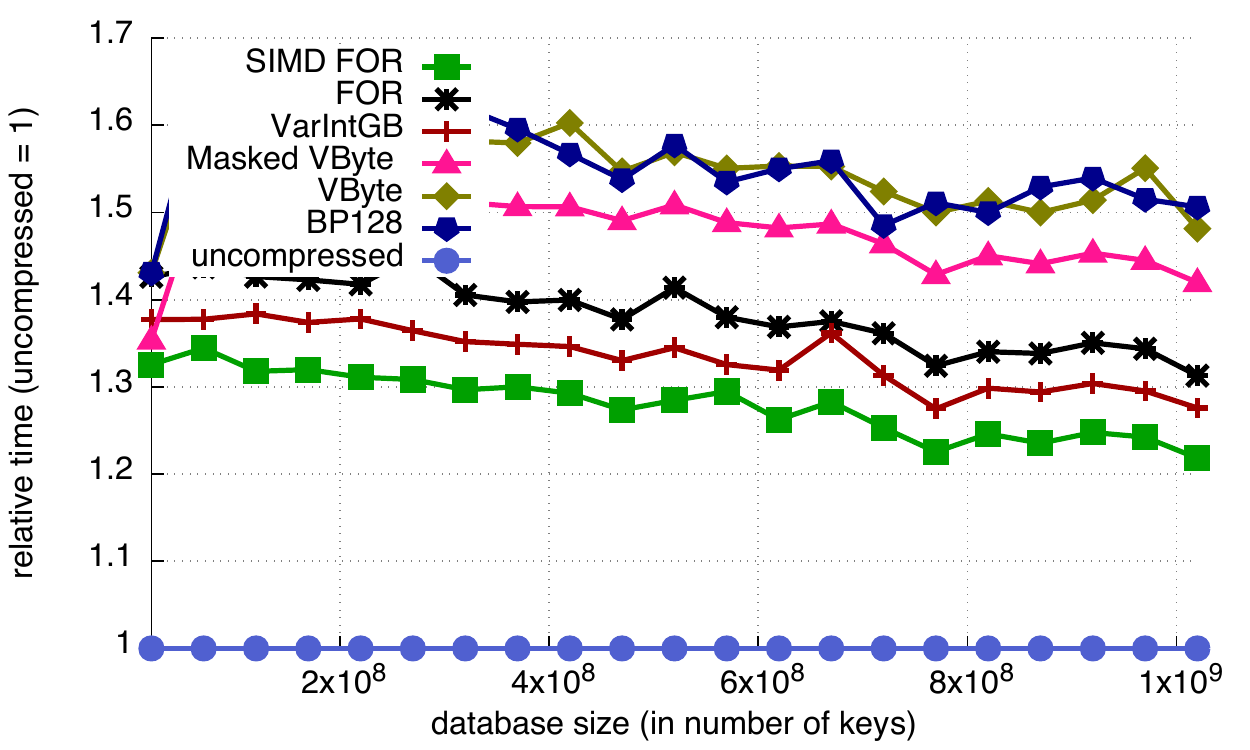} }
\subfloat[Relative cursor timings \label{fig:relcursor}]{\centering\includegraphics[width=0.99\columnwidth]{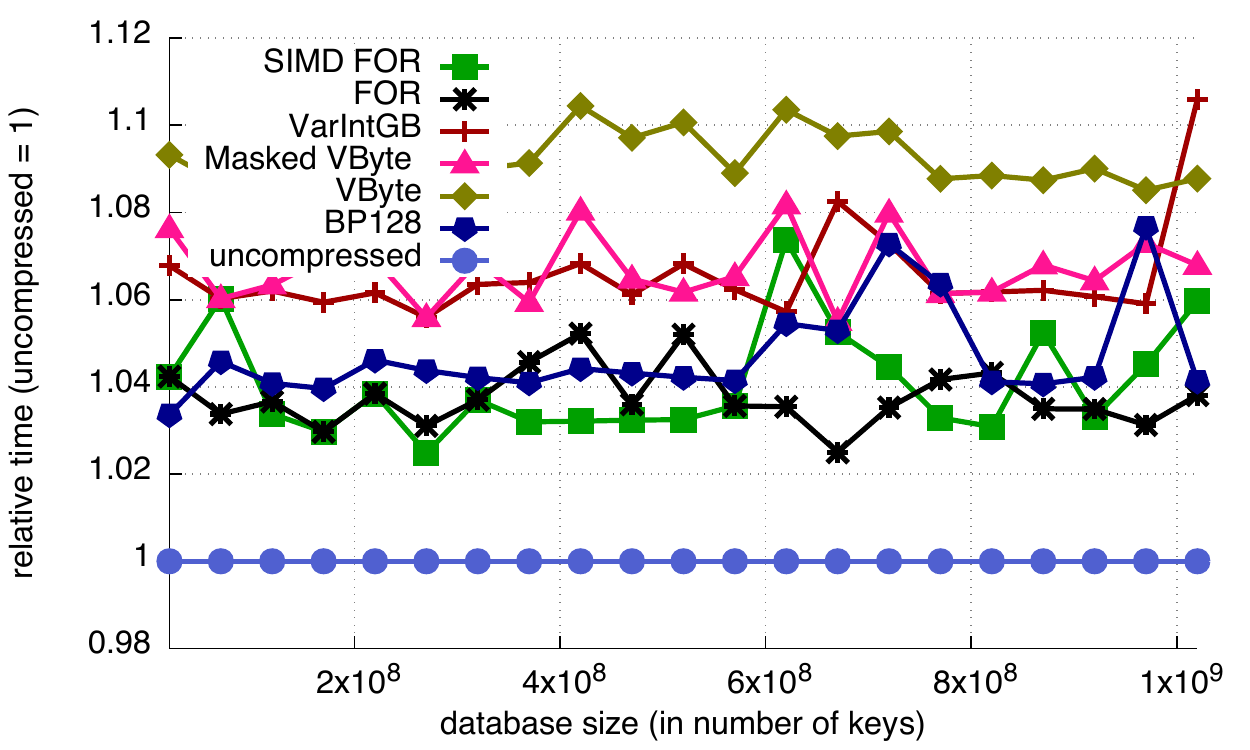}}
\\
\subfloat[Relative SUM timings \label{fig:relsum}]{\centering\includegraphics[width=0.99\columnwidth]{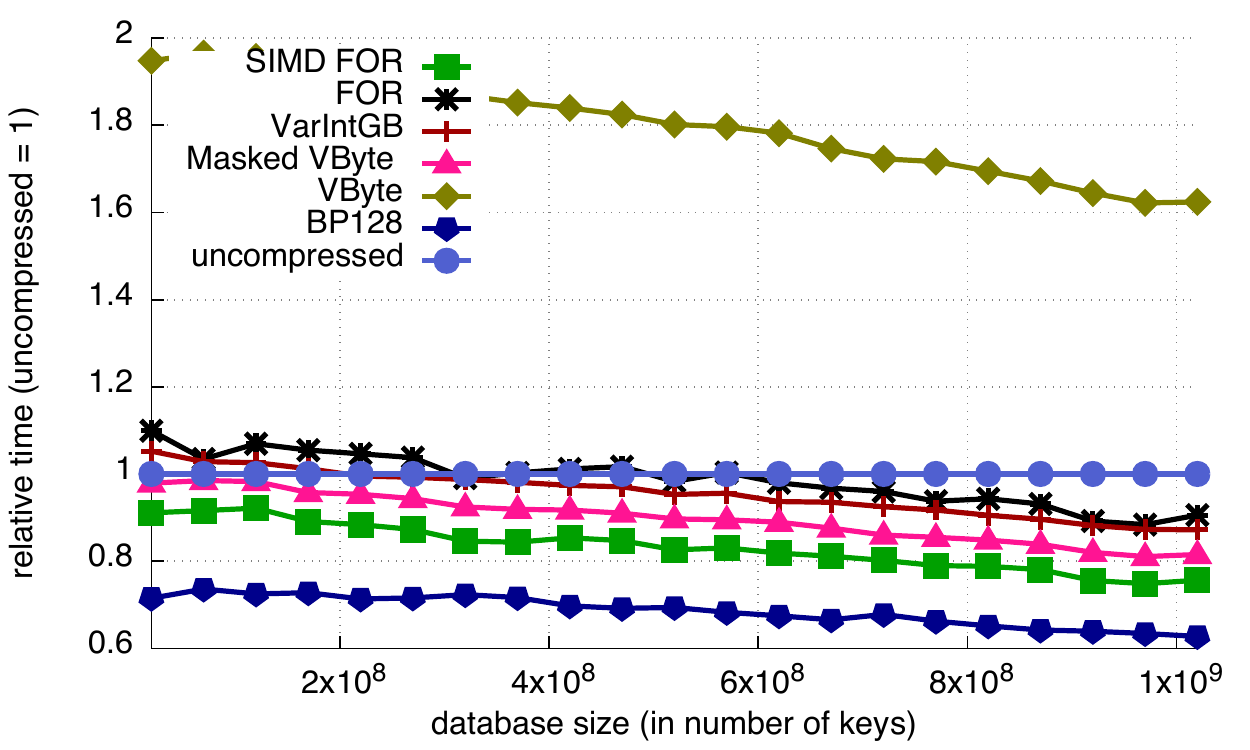}}
\subfloat[Relative insert timings \label{fig:relins}]{\centering\includegraphics[width=0.99\columnwidth]{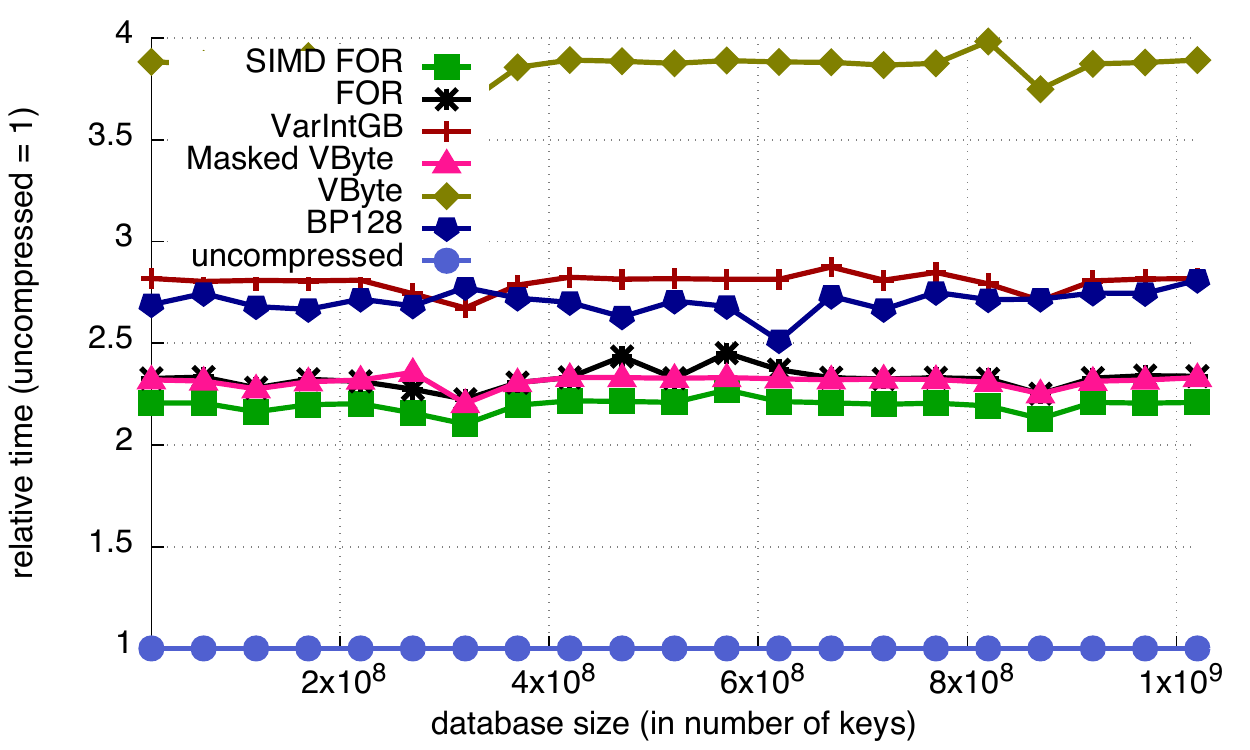}}
\\
\caption{Relative  timings in Upscaledb (ClusterData). \label{fig:relresults} }
\end{figure*}

We execute four operations:
\begin{description}
\item[Look-up]
This benchmark opens an existing database and performs point-lookups of all
inserted keys. Each lookup requires a B+-tree traversal to the leaf node.
The node then performs a linear search through the block index and locates
the block which stores the requested key. The codec then searches the block
for the key.

We implemented search functions for all codecs directly on compressed data.
FOR and SIMD~FOR do not use differential compression and therefore perform a
binary search directly on the compressed data. All other codecs use linear
search because they need to rebuild the original value during the search.

The benchmarks show that integer compression does not cause a significant
performance hit for lookup operations. Indeed, Fig.~\ref{fig:rellookup} shows that the penalty for using compressed keys is about \SI{50}{\percent}. We get the best results with SIMD~FOR, VarIntGB and FOR (a penalty ranging from  \SI{20}{\percent} to \SI{40}{\percent}) and the worst results with VByte and BP128 (with a penalty of up to \SI{60}{\percent}). The good results from VarIntGB in this case are consistent with our microbenchmarks (see \S~\ref{sec:micro} and Fig.~\ref{fig:search}). BP128 is slightly penalized in this case because it is able to store more keys per leaf node due to its better compression: searching in a node containing more keys takes longer on average, everything else being equal. Though VByte and Masked~VByte have the same underlying format, Masked~VByte is noticeably faster though not as fast as FOR and SIMD~FOR.

\item[Cursor]
This benchmark opens an existing database and creates a cursor to traverse from
the first to the last key. To position itself on the first key, the cursor
traverses down the B+-tree at the left-most path down to the leaf,
then visits each leaf. Since all leaf nodes are linked, no further traversal
is required.

The cursor attaches itself to a leaf node, and stores the current position
in the leaf. When the cursor is moved to the next key, this position is
incremented. If the last key in the leaf is reached, the cursor loads the
sibling of the leaf, attaches itself to the sibling and resets its position
to 0.

The cursor then retrieves the key at its current position.
If the \texttt{KeyList} is uncompressed then the key is accessed with $O(1)$.
A compressed \texttt{KeyList} first traverses the list of
block descriptors, accumulating each block's number of keys till it finds
the block which contains the requested key.
In our original implementation, the cursor then used a \texttt{select} method
to retrieve the key directly from the compressed block. But since
cursors are usually used for sequential access, and therefore frequently
access the same block, we decided to decode the block and cache the decoded
values. This causes additional latency when a block is accessed
initially, but all following accesses can be served with high throughput.
Indeed, our tests showed a significant performance improvement compared to
the previous implementation based on \texttt{select}.

The final results are presented in Fig.~\ref{fig:relcursor}. Compared to an uncompressed database, all codecs except VByte show a penalty of less than about \SI{8}{\percent}. VByte does slightly worse with a penalty sometimes exceeding slightly \SI{11}{\percent}. Again, though they use the same underlying format, Masked~VByte is noticeably faster than VByte.

\item[SUM]

This benchmark performs a ``SUM'' operation on all keys. It is equivalent to
a \texttt{SELECT SUM(column)} operation of a SQL database, where the specified
column is an index of unique 32-bit integer keys. For such operations,
Upscaledb does not use a cursor to traverse the B+-tree, but performs the
operation directly on the B+-tree's data, without copying the keys into the
application's memory.

If compression is disabled, the \texttt{KeyList} stores all keys in an array
of type \texttt{uint32\_t[]}. The ``SUM'' operation  sums all keys in
that array. If compression is enabled, the \texttt{KeyList} traverses each
compressed block, uncompresses it into temporary memory (in L1 cache) and sums all keys
of that memory.

The benchmark results in Fig.~\ref{fig:relsum} show the SIMD accelerated
BP128 and SIMD~FOR as the clear winners. The compressed databases are even faster than an uncompressed database---with gains reaching \SI{40}{\percent} for BP128.

SUM performance is impacted by database 
size: the bigger the database, the more compression is beneficial, with BP128 and SIMD~FOR 
offering the best performance. Only Masked~VByte, BP128 and SIMD~FOR are superior to the uncompressed database 
on the entire test range. VarIntGB and FOR also help the speed for large databases while VByte fails to catch up to the uncompressed database in the scope of our test. 

We take this query as a representative of 
analytic queries where much of the data must be accessed (the query has low selectivity). In such cases,
we expect compression to be particularly useful as
it reduces data access costs. 
Upscaledb supports several such queries such as 
COUNT, COUNT DISTINCT, COUNT\_IF, COUNT\_DISTINCT\_IF, AVERAGE, \ldots 

Fig.~\ref{fig:newrelresults} illustrates our results with a more advanced query (``AVERAGE(key) WHERE key $>$ MAX(keys) / 2''). It shows that such queries can be accelerated by  the fast compression offered by SIMD~FOR and BP128.

\begin{figure}[tbh]
\centering\includegraphics[width=0.99\columnwidth]{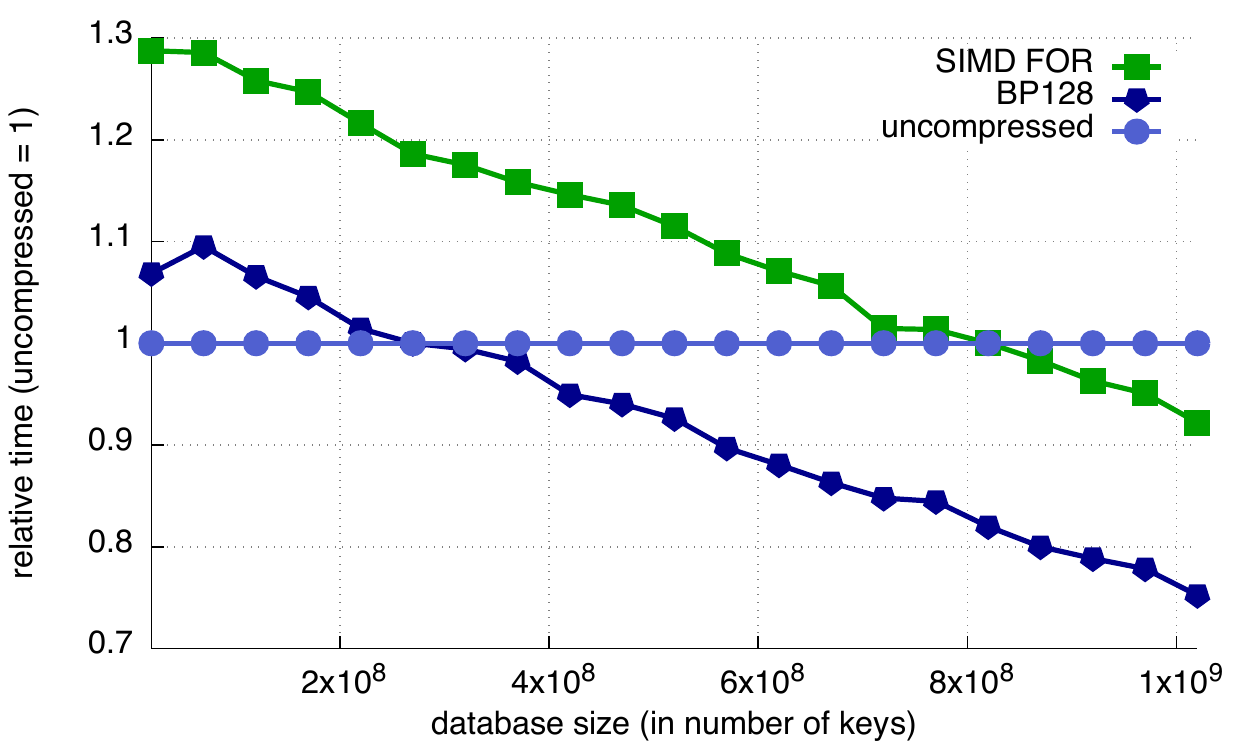}
\caption{Relative  timings in Upscaledb for the query ``AVERAGE(key) WHERE key $>$ MAX(keys) / 2'' over the ClusterData dataset. \label{fig:newrelresults} }
\end{figure}

\item[Insert]
This benchmark creates a new database for 32-bit integer keys and inserts 
various numbers of keys. We should expect a compressed database to be slower for such applications, as insertions may require complete recompression of some blocks---in the worst case.

Fig.~\ref{fig:relins} shows that among the compressed formats, the best insertion performance is offered by the FOR, SIMD~FOR and Masked~VByte codecs, followed by BP128 and VarIntGB\@. VByte is slower than all other codecs.
If one uses FOR, SIMD~FOR and Masked~VByte, insertions in a compressed database are  only $2.5\times$ slower than insertions in an uncompressed database.
\end{description}

\subsubsection{Setting the block size}
\label{sec:blocksize}
\begin{figure*}[tbh]
\centering
\subfloat[Database sizes]{\centering\includegraphics[width=0.99\columnwidth]{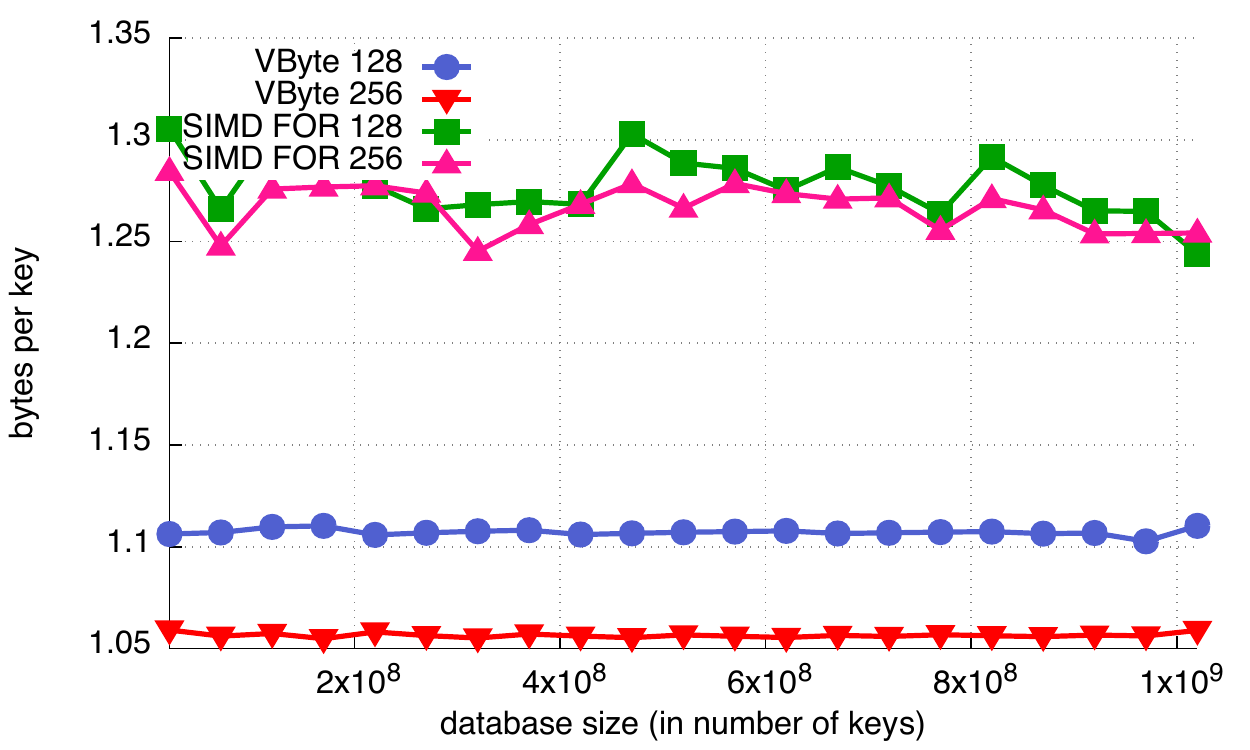}}
\subfloat[SUM timings]{\centering\includegraphics[width=0.99\columnwidth]{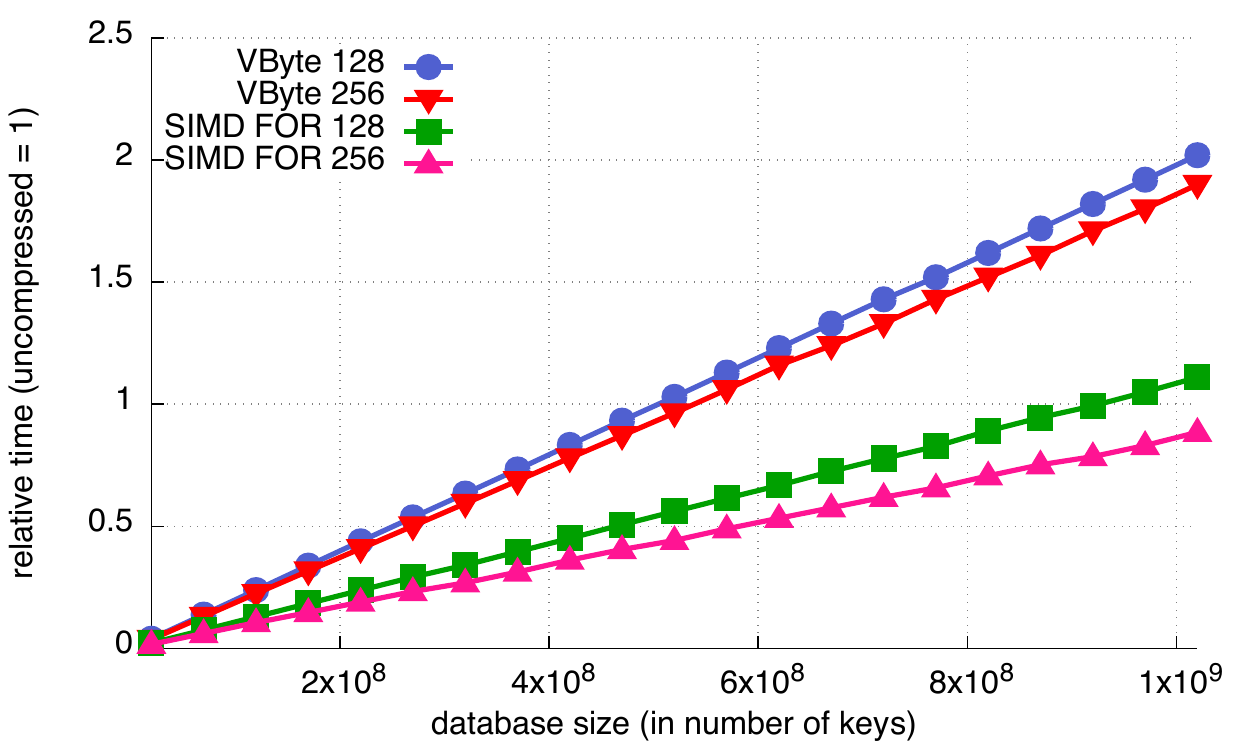}}
\caption{Experimental results for two block sizes (128, 256)  over ClusterData  in Upscaledb. \label{fig:datim} }
\end{figure*}

Fig.~\ref{fig:datim} presents the same results for two possible block sizes (128, 256). We experimented with 
a wide range of block sizes, but only report on these two choices for simplicity. We see that the performance with a larger block size (256) is slightly better and the overall size smaller. The FOR and SIMD~FOR codecs  benefit substantially from larger block sizes (compared to other codecs) because they rely on a binary search to locate values in the compressed stream: the benefits of a binary search versus a sequential search grow with the size of the blocks.   
 This justifies our design choice of opting for large block sizes (256) for all but one codec. Exceptionally, for BP128, we prefer the smaller block size (128).

\section{Related Work}
\label{sec:related}

There has been much research dedicated to improving the performance of tree data structures. For example, cache-conscious trees can reduce the number of cache misses to improve performance~\cite{4041597,cst}.
In particular, Lee et al.~\cite{cst} propose the Cache Sensitive T-Trees (CST-Trees).

The application of SIMD instructions to accelerate B+-tree operations (without compression) is reviewed by Zhou and Ross~\cite{Zhou:2002:IDO:564691.564709}.
Willhalm et al.~\cite{Willhalm:2009:SUF:1687627.1687671} describe how to scan quickly column stores using SIMD instructions.
Schlegel et al.~\cite{Schlegel:2009:KSM:1565694.1565705} show how to accelerate K-ary search on modern processors.
Raman et al.~\cite{Raman:2013:DBA:2536222.2536233} describe the IBM DB2 column store that makes extensive use of SIMD instructions and compression.

Compression in databases has a long history~\cite{Westmann:2000:IPC:362084.362137}.
Compression techniques such as run-length encoding and differential coding are common, for example, in column-oriented databases~\cite{Stonebraker:2005:CCD:1083592.1083658}.
IBM DB2 compresses integer keys using variable-byte compression and differential coding~\cite{db2luw2009}. Graefe~\cite{Graefe:2007:ECS:1276301.1276302}
 describes the compression opportunity when  keys are consecutive.

Jin and Chung improve the CST-Trees by using FOR compression~\cite{5593131}.
Similarly, Kim et al.~\cite{Kim:2010:FFA:1807167.1807206} propose a SIMD-accelerated in-memory tree index (FAST) where they use FOR\@.
In related work, Yamamuro et al.~\cite{Yamamuro:2012:VVC:2247596.2247643} propose the VAST-Tree: it improves upon FAST in several ways. In particular, it offers better compression ratios of the keys than FAST by using differential coding and the PFOR compression scheme~\cite{1617427}. Though fast, PFOR does not exploit SIMD instructions:  Lemire and Boytsov~\cite{LemireBoytsov2013decoding} found that SIMD-accelerated binary packing (i.e., BP128) could be 2 to 3 times faster with little difference in the compression ratios. 

Random access in differentially-coded compressed arrays is 
often made possible with auxiliary data structures that allow skipping~\cite{Transier:2010:EBA:1877766.1877768,moffat1996self}.  However, there are alternatives to differential coding that offer more convenient random access. Claude et al.~\cite{Claude2014272} propose differentially encoded search trees; Teuhola~\cite{Teuhola2011742} adapts interpolative coding~\cite{Moffat2000binary} so that it can support logarithmic search.
Brisaboa et al.~\cite{Brisaboa2013392} modify variable-byte encoding to create Directly Addressable Codes (DACs)---so that one can have access to individual coded value in constant time using rank/select dictionaries.
This strategy is applied
to other compression schemes
by K\"ulekci~\cite{6824444}.
 There has also been much interest
in variations on the Elias-Fano representation~\cite{Vigna2013,Ottaviano:2014:PEI:2600428.2609615}, as it can provide good compression and fast random access to the encoded values. Other techniques such as wavelet trees~\cite{Navarro20142} or bitmap indexes~\cite{SPE:SPE2402,Culpepper:2010:ESI:1877766.1877767,kane2014,SPE:SPE2326} can also be used for similar purposes.


Our work should be applicable to other B-trees
and related data structures, i.e.,  Log-Structured Merge-Trees~\cite{O'Neil:1996:LM:230823.230826,Chang:2008:BDS:1365815.1365816} (LSM).

\section{Conclusion}

We have shown that fast key compression could improve the single-threaded performance
of a key-value store---if these compression techniques 
are accelerated with SIMD instructions. One of our best
performing codec (BP128) has the property that the removal
of a key may (slightly) increase the storage requirement: 
something that the IBM DB2 design team specifically excluded.
We have presented a practical B+-tree
implementation that supports  this case where the deletion of a key 
may increase the storage.

We get the best performance for SIMD codecs (BP128 and SIMD~FOR). Unlike other codecs, 
they show an ability to noticeably improve query performance in all our tests (from small
to large databases) on the analytic (SUM) benchmark. Naturally, these gains come with reduced storage usage. As we expected,
there is a downside to compression: slower insertion operations.  
However, for analytic 
applications where insertions are infrequent, this downside may be inconsequential.
The choice between BP128 and SIMD~FOR is a trade-off between superior compression (BP128) and superior random lookup speed (SIMD~FOR). Indeed, SIMD~FOR has superior lookup performance because it supports a binary search directly on the compressed data. However, BP128 has superior compression due to its reliance on differential coding. Our experiments show that BP128 has better performance for low selectivity queries.
 Our results also show that if we are to use the standard VByte format, then a SIMD-accelerated decoder (Masked~VByte) can accelerate queries without requiring any change to the database format.

Our results suggest also that it is beneficial to program common analytic functions (e.g., SUM, COUNT, COUNT DISTINCT, AVERAGE) so that they work directly on buffered data in CPU cache, bypassing explicit cursor handling. Further work could
quantify the benefits of implementing these functions so that they operate directly on compressed data.

The importance of SIMD instructions for performance 
is likely to grow. Already, some processors support wider registers (e.g., 256~bits for recent Intel processors using AVX2 and 512~bits for upcoming processors using AVX-512). From an engineering perspective, 
it seems easier to design processors that operate on more values during each cycle (wider processors) than processors that run at a higher frequency.
Thus it is probably wise to invest in data structures
that are best able to benefit from SIMD instructions.

The SIMD accelerated BP128 also offers the best compression ratio, especially
for dense key ranges like auto-incremented primary keys or dense time stamps.
BP128 compresses data by an order of magnitude,
reducing pressure on memory resources.

We limited our work to the compression of 32-bit integer keys. This is a common case 
well worth optimizing. However, many of the other compression techniques developed for
B+-trees could be revisited in light of the new hardware capabilities.

\section{Acknowledgments}
D.~Lemire acknowledges support from the Natural Sciences and Engineering Research 
Council of Canada (NSERC) from grant number 26143.

\section{References}

\balance

\bibliographystyle{abbrv}
\bibliography{bib/acmtitles.bib,bib/shorttitlesiso.bib,bib/lemur.bib}  

\begin{thebibliography}{10}

\bibitem{Anh:2010:ICU:1712666.1712668}
V.~N. Anh and A.~Moffat.
\newblock Index compression using 64-bit words.
\newblock {\em Softw. Pract. Exp.}, 40(2):131--147, 2010.

\bibitem{Bayer:1972:OML:2697402.2697450}
R.~Bayer and E.~M. Mccreight.
\newblock Organization and maintenance of large ordered indexes.
\newblock {\em Acta Inf.}, 1(3):173--189, Sept. 1972.

\bibitem{db2luw2009}
B.~Bhattacharjee, L.~Lim, T.~Malkemus, G.~Mihaila, K.~Ross, S.~Lau,
  C.~McArthur, Z.~Toth, and R.~Sherkat.
\newblock Efficient index compression in {DB2 LUW}.
\newblock {\em Proc. VLDB Endow.}, 2(2):1462--1473, 2009.

\bibitem{1559877}
C.~Binnig, S.~Hildenbrand, and F.~F\"{a}rber.
\newblock Dictionary-based order-preserving string compression for main memory
  column stores.
\newblock In {\em SIGMOD 2009}, pages 283--296, New York, NY, USA, 2009. ACM.

\bibitem{Brisaboa2013392}
N.~R. Brisaboa, S.~Ladra, and G.~Navarro.
\newblock {DACs}: Bringing direct access to variable-length codes.
\newblock {\em Inform. Process. Manag.}, 49(1):392 -- 404, 2013.

\bibitem{buttcher2010information}
S.~B{\"u}ttcher, C.~Clarke, and G.~Cormack.
\newblock {\em Information retrieval: Implementing and evaluating search
  engines}.
\newblock The MIT Press, Cambridge, Massachusetts, 2010.

\bibitem{buttcher}
S.~B\"uttcher and C.~L. Clarke.
\newblock Hybrid index maintenance for contiguous inverted lists.
\newblock {\em Information Retrieval}, 11(3):175--207, 2008.

\bibitem{Chang:2008:BDS:1365815.1365816}
F.~Chang, J.~Dean, S.~Ghemawat, W.~C. Hsieh, D.~A. Wallach, M.~Burrows,
  T.~Chandra, A.~Fikes, and R.~E. Gruber.
\newblock {BigTable}: A distributed storage system for structured data.
\newblock {\em ACM Trans. Comput. Syst.}, 26(2):4:1--4:26, June 2008.

\bibitem{Claude2014272}
F.~Claude, P.~K. Nicholson, and D.~Seco.
\newblock On the compression of search trees.
\newblock {\em Inform. Process. Manag.}, 50(2):272--283, 2014.

\bibitem{Comer:1979:UB:356770.356776}
D.~Comer.
\newblock Ubiquitous b-tree.
\newblock {\em ACM Comput. Surv.}, 11(2):121--137, June 1979.

\bibitem{Culpepper:2010:ESI:1877766.1877767}
J.~S. Culpepper and A.~Moffat.
\newblock Efficient set intersection for inverted indexing.
\newblock {\em ACM Trans. Inf. Syst.}, 29(1):1:1--1:25, Dec. 2010.

\bibitem{DeanOfficialplusslides:2009:CBL:1498759.1498761}
J.~Dean.
\newblock Challenges in building large-scale information retrieval systems:
  invited talk.
\newblock In {\em Proceedings of the Second ACM International Conference on Web
  Search and Data Mining}, WSDM '09, pages 1--1, New York, NY, USA, 2009. ACM.
\newblock Author's slides:
  \url{http://static.googleusercontent.com/external_content/untrusted_dlcp/research.google.com/en/us/people/jeff/WSDM09-keynote.pdf}
  [Last checked Nov. 2015.].

\bibitem{655800}
J.~Goldstein, R.~Ramakrishnan, and U.~Shaft.
\newblock Compressing relations and indexes.
\newblock In {\em Proceedings of the Fourteenth International Conference on
  Data Engineering}, ICDE '98, pages 370--379, Washington, DC, USA, 1998. IEEE
  Computer Society.

\bibitem{Graefe:2007:ECS:1276301.1276302}
G.~Graefe.
\newblock Efficient columnar storage in {B}-trees.
\newblock {\em SIGMOD Rec.}, 36(1):3--6, Mar. 2007.

\bibitem{Guibas:1978:DFB:1382432.1382565}
L.~J. Guibas and R.~Sedgewick.
\newblock A dichromatic framework for balanced trees.
\newblock In {\em Proceedings of the 19th Annual Symposium on Foundations of
  Computer Science}, SFCS '78, pages 8--21, Washington, DC, USA, 1978. IEEE
  Computer Society.

\bibitem{5593131}
R.~Jin and T.-S. Chung.
\newblock Node compression techniques based on {Cache-Sensitive B+-Tree}.
\newblock In {\em Computer and Information Science (ICIS), 2010 IEEE/ACIS 9th
  International Conference on}, pages 133--138, Aug 2010.

\bibitem{kane2014}
A.~Kane and F.~Tompa.
\newblock Skewed partial bitvectors for list intersection.
\newblock In {\em Proceedings of the 37th annual international ACM SIGIR
  conference on Research and development in information retrieval}, pages
  263--272. ACM, 2014.

\bibitem{Kim:2010:FFA:1807167.1807206}
C.~Kim, J.~Chhugani, N.~Satish, E.~Sedlar, A.~D. Nguyen, T.~Kaldewey, V.~W.
  Lee, S.~A. Brandt, and P.~Dubey.
\newblock {FAST}: Fast architecture sensitive tree search on modern cpus and
  gpus.
\newblock In {\em Proceedings of the 2010 ACM SIGMOD International Conference
  on Management of Data}, SIGMOD '10, pages 339--350, New York, NY, USA, 2010.
  ACM.

\bibitem{6824444}
M.~Kulekci.
\newblock Enhanced variable-length codes: Improved compression with efficient
  random access.
\newblock In {\em Data Compression Conference (DCC), 2014}, pages 362--371,
  March 2014.

\bibitem{cst}
I.-h. Lee, J.~Shim, S.-g. Lee, and J.~Chun.
\newblock {CST-Trees}: {Cache Sensitive T-Trees}.
\newblock In R.~Kotagiri, P.~Krishna, M.~Mohania, and E.~Nantajeewarawat,
  editors, {\em Advances in Databases: Concepts, Systems and Applications},
  volume 4443 of {\em Lecture Notes in Computer Science}, pages 398--409.
  Springer Berlin Heidelberg, 2007.

\bibitem{LemireBoytsov2013decoding}
D.~Lemire and L.~Boytsov.
\newblock Decoding billions of integers per second through vectorization.
\newblock {\em Softw. Pract. Exp.}, 45(1):1--29, 2015.

\bibitem{SPE:SPE2326}
D.~Lemire, L.~Boytsov, and N.~Kurz.
\newblock {SIMD} compression and the intersection of sorted integers.
\newblock {\em Softw. Pract. Exp.}, 46(6):723--749, 2016.

\bibitem{SPE:SPE2402}
D.~Lemire, G.~Ssi-Yan-Kai, and O.~Kaser.
\newblock Consistently faster and smaller compressed bitmaps with roaring.
\newblock {\em Software: Practice and Experience}, 46(11):1547--1569, 2016.

\bibitem{4041597}
G.~Mihaila and I.~Stanoi.
\newblock A tree for all seasons.
\newblock In {\em Database Engineering and Applications Symposium, 2006. IDEAS
  '06. 10th International}, pages 3--10, Dec 2006.

\bibitem{Moffat2000binary}
A.~Moffat and L.~Stuiver.
\newblock Binary interpolative coding for effective index compression.
\newblock {\em Inform. Retrieval}, 3(1):25--47, 2000.

\bibitem{moffat1996self}
A.~Moffat and J.~Zobel.
\newblock Self-indexing inverted files for fast text retrieval.
\newblock {\em ACM Trans. Inf. Syst.}, 14(4):349--379, 1996.

\bibitem{Navarro20142}
G.~Navarro.
\newblock Wavelet trees for all.
\newblock {\em Journal of Discrete Algorithms}, 25(0):2 -- 20, 2014.
\newblock 23rd Annual Symposium on Combinatorial Pattern Matching.

\bibitem{Olson:1999:BD:1268708.1268751}
M.~A. Olson, K.~Bostic, and M.~Seltzer.
\newblock Berkeley db.
\newblock In {\em Proceedings of the Annual Conference on USENIX Annual
  Technical Conference}, ATEC '99, pages 43--43, Berkeley, CA, USA, 1999.
  USENIX Association.

\bibitem{O'Neil:1996:LM:230823.230826}
P.~O'Neil, E.~Cheng, D.~Gawlick, and E.~O'Neil.
\newblock The log-structured merge-tree (lsm-tree).
\newblock {\em Acta Inf.}, 33(4):351--385, June 1996.

\bibitem{Ottaviano:2014:PEI:2600428.2609615}
G.~Ottaviano and R.~Venturini.
\newblock Partitioned elias-fano indexes.
\newblock In {\em Proceedings of the 37th International ACM SIGIR Conference on
  Research \&\#38; Development in Information Retrieval}, SIGIR '14, pages
  273--282, New York, NY, USA, 2014. ACM.

\bibitem{maskedvbyte}
J.~Plaisance, N.~Kurz, and D.~Lemire.
\newblock {Vectorized VByte Decoding}.
\newblock In {\em Proceedings of the first International Symposium on Web
  Algorithms}, iSWAG '15, 2015.
\newblock Available from \url{http://arxiv.org/abs/1503.07387}.

\bibitem{Raman:2013:DBA:2536222.2536233}
V.~Raman, G.~Attaluri, R.~Barber, N.~Chainani, D.~Kalmuk, V.~KulandaiSamy,
  J.~Leenstra, S.~Lightstone, S.~Liu, G.~M. Lohman, T.~Malkemus, R.~Mueller,
  I.~Pandis, B.~Schiefer, D.~Sharpe, R.~Sidle, A.~Storm, and L.~Zhang.
\newblock Db2 with blu acceleration: So much more than just a column store.
\newblock {\em Proc. VLDB Endow.}, 6(11):1080--1091, Aug. 2013.

\bibitem{Schlegel:2009:KSM:1565694.1565705}
B.~Schlegel, R.~Gemulla, and W.~Lehner.
\newblock $k$-ary search on modern processors.
\newblock In {\em Proceedings of the Fifth International Workshop on Data
  Management on New Hardware}, DaMoN '09, pages 52--60, New York, NY, USA,
  2009. ACM.

\bibitem{Stepanov:2011:SDP:2063576.2063627}
A.~A. Stepanov, A.~R. Gangolli, D.~E. Rose, R.~J. Ernst, and P.~S. Oberoi.
\newblock {SIMD}-based decoding of posting lists.
\newblock In {\em Proceedings of the 20th ACM International Conference on
  Information and Knowledge Management}, CIKM '11, pages 317--326, New York,
  NY, USA, 2011. ACM.

\bibitem{Stonebraker:2005:CCD:1083592.1083658}
M.~Stonebraker, D.~J. Abadi, A.~Batkin, X.~Chen, M.~Cherniack, M.~Ferreira,
  E.~Lau, A.~Lin, S.~Madden, E.~O'Neil, P.~O'Neil, A.~Rasin, N.~Tran, and
  S.~Zdonik.
\newblock {C-store}: A column-oriented dbms.
\newblock In {\em Proceedings of the 31st International Conference on Very
  Large Data Bases}, VLDB '05, pages 553--564. VLDB Endowment, 2005.

\bibitem{Teuhola2011742}
J.~Teuhola.
\newblock Interpolative coding of integer sequences supporting log-time random
  access.
\newblock {\em Inform. Process. Manag.}, 47(5):742--761, 2011.
\newblock Managing and Mining Multilingual Documents.

\bibitem{Transier:2010:EBA:1877766.1877768}
F.~Transier and P.~Sanders.
\newblock Engineering basic algorithms of an in-memory text search engine.
\newblock {\em ACM Trans. Inf. Syst.}, 29(1):2:1--2:37, Dec. 2010.

\bibitem{Vigna2013}
S.~Vigna.
\newblock Quasi-succinct indices.
\newblock In {\em Proceedings of the Sixth ACM International Conference on Web
  Search and Data Mining}, WSDM '13, pages 83--92, New York, NY, USA, 2013.
  ACM.

\bibitem{Westmann:2000:IPC:362084.362137}
T.~Westmann, D.~Kossmann, S.~Helmer, and G.~Moerkotte.
\newblock The implementation and performance of compressed databases.
\newblock {\em SIGMOD Rec.}, 29(3):55--67, Sept. 2000.

\bibitem{Willhalm:2009:SUF:1687627.1687671}
T.~Willhalm, N.~Popovici, Y.~Boshmaf, H.~Plattner, A.~Zeier, and J.~Schaffner.
\newblock {SIMD}-scan: ultra fast in-memory table scan using on-chip vector
  processing units.
\newblock {\em Proc. VLDB Endow.}, 2(1):385--394, Aug. 2009.

\bibitem{williams1999compressing}
H.~E. Williams and J.~Zobel.
\newblock Compressing integers for fast file access.
\newblock {\em The Computer Journal}, 42(3):193--201, 1999.

\bibitem{Yamamuro:2012:VVC:2247596.2247643}
T.~Yamamuro, M.~Onizuka, T.~Hitaka, and M.~Yamamuro.
\newblock {VAST-Tree}: A vector-advanced and compressed structure for massive
  data tree traversal.
\newblock In {\em Proceedings of the 15th International Conference on Extending
  Database Technology}, EDBT '12, pages 396--407, New York, NY, USA, 2012. ACM.

\bibitem{Zhao:2015:GSA:2737814.2735629}
W.~X. Zhao, X.~Zhang, D.~Lemire, D.~Shan, J.-Y. Nie, H.~Yan, and J.-R. Wen.
\newblock A general {SIMD}-based approach to accelerating compression
  algorithms.
\newblock {\em ACM Trans. Inf. Syst.}, 33(3):15:1--15:28, Mar. 2015.

\bibitem{Zhou:2002:IDO:564691.564709}
J.~Zhou and K.~A. Ross.
\newblock Implementing database operations using {SIMD} instructions.
\newblock In {\em Proceedings of the 2002 ACM SIGMOD International Conference
  on Management of Data}, SIGMOD '02, pages 145--156, New York, NY, USA, 2002.
  ACM.

\bibitem{1617427}
M.~Zukowski, S.~Heman, N.~Nes, and P.~Boncz.
\newblock Super-scalar {RAM-CPU} cache compression.
\newblock In {\em Proceedings of the 22nd International Conference on Data
  Engineering}, ICDE '06, pages 59--71, Washington, DC, USA, 2006. IEEE
  Computer Society.

\end{thebibliography}
\appendix

\section{Insertions and Deletions in B+-trees}
\label{appendix:insertionanddeletions}

B+-tree~\cite{Comer:1979:UB:356770.356776} can be considered
textbook material. Nevertheless, for completeness, we briefly
review insertions and deletions in B+-trees.
Non-root, non-leaf nodes can accommodate between $b$ and $2b$~keys
whereas  leaf nodes accommodate between
$b$ and $2b-1$~keys. Nodes are split or merged to maintain
the number of keys
in these ranges.

Insertion in a B+-tree
works generally as a two-step process: we first
go down the tree to find the right node, and then
we potentially split the nodes from the bottom up.
That is, when inserting, we go down the tree to find the leaf node
where the insertion is to happen.
If the leaf can accommodate another key (its cardinality is less than $2b-1$),
it is inserted and the process terminates.
When the node is already full, it is split:
we divide the node into two nodes containing $b$~keys.
One of the new nodes contains the smallest $b$~keys
whereas the other one contains the largest $b$~keys.
The smallest value of the latter node is copied and added
to the parent node as a separator. If needed,
the parent node is split too, and the process recurses,
possibly all the way to the root. If the root needs
to be split, then the height of the tree is effectively increased
by one.

Deletion proceeds similarly at first. We find
the  appropriate leaf node where the key resides.
If deleting the key would leave at least $b$~keys in the
node, we proceed and the process terminates. If there are
two few values, we can examine a neighboring leaf node, having the same parent
if possible.
If it has more than $b$~keys, it suffices to borrow one of
the keys. If taken together, the two leaf nodes have less
than $2b$~keys, they need to be merged. Their merger implies
at least the removal of the separator key in the parent node, and the
process may recurse up to the root, possibly decreasing the 
height of the tree.

\end{document}